\documentclass[a4paper,11pt]{article}
\pdfoutput=1 
\usepackage{jheppub} 
                     
\usepackage{booktabs,makecell, bm}
\usepackage[font=small]{caption}
\usepackage{framed}
\usepackage{hyperref}
\usepackage[export]{adjustbox}
\usepackage{amsmath}

\usepackage{url}
\usepackage[numbers]{natbib}
\usepackage{physics}
\usepackage{subcaption}

\title{Large charge operators at large spin from relativistically rotating vortices}

\author[a]{Jaehyeok Choi,}
\author[b]{Eunwoo Lee}

\affiliation[a]{Department of Physics and Astronomy \& Center for
Theoretical Physics,\\
Seoul National University, Seoul 08826, Korea.}

\affiliation[b]{Department of Theoretical Physics, \\ Tata Institute
	of Fundamental Research, Homi Bhabha Rd, Mumbai 400005, India}
\emailAdd{zaddere7jp@snu.ac.kr}
\emailAdd{eunwoo.lee@tifr.res.in}

\abstract{We study the ground states of CFTs with a global \(U(1)\) symmetry on \(\mathbb{R}\times S^2\) in the regime of large charge $Q$ and large angular momentum $J$, using large charge EFT. We find that in the range \(Q \ll J \ll Q^2\), the ground state solution is a superfluid densely populated with vortices rotating at a constant angular velocity $\Omega$. This is a relativistic generalization of the known (non-relativistic) rigid rotation phase, which corresponds to the small $\Omega$ limit of our solution.
In the regime $Q^{3/2}\ll J\ll Q^2$, our solution achieves lower energy than previously identified states.
In this regime, most of the vortices move near the speed of light, and we obtain the chiral fluctuation modes propagating at the speed of light.
Interestingly, we find that our ground state can be interpreted as a zero temperature charged normal fluid rotating at a constant angular velocity $\Omega$.
We rederive this solution purely from the fluid dynamics. 
Based on the (already established) applicability of fluid description to large non-supersymmetric extremal AdS black holes, we find that the boundary stress tensor and $U(1)$ current of extremal AdS Kerr-Newman black hole align with those of our solution.
}

\makeatletter
\gdef\@fpheader{}
\makeatother

\begin{document} 

\maketitle

\section{Introduction}
Finding the ground state of a conformal field theory (CFT) on $\mathbb{R}\times S^{d-1}$ at fixed quantum numbers is an intriguing problem as it corresponds to determining the local operator with the smallest scaling dimension. Significant progress has been made in determining the relationship between the ground state energy and conserved quantities, at large charge \cite{Hellerman:2015nra,Alvarez-Gaume:2016vff, Loukas:2016ckj, Loukas:2017lof, Hellerman:2017efx, Hellerman:2017veg, Hellerman:2017sur, Monin:2016jmo, Jafferis:2017zna, Bourget:2018obm, Hellerman:2018xpi, Cuomo:2019ejv,Badel:2019khk,Cuomo:2023mxg,Badel:2022fya,Dondi:2024vua} and large angular momentum \cite{Alday:2007mf,Komargodski:2012ek,Fitzpatrick:2012yx,Fitzpatrick:2014vua,Kaviraj:2015cxa,Alday:2015eya,Simmons-Duffin:2016wlq,Gopakumar:2016cpb,Caron-Huot:2017vep,Fardelli:2024heb}.

For large charge, by now the canonical approach is to study an effective field theory (EFT) of the CFT. By assuming a large charge, we introduce a new scale in the CFT.
In the Wilsonian effective action, the hierarchy of scales is given as follows:
\begin{align}
    \frac{1}{R} \ll \Lambda \ll {\rho}^{\frac{1}{d-1}} \propto \frac{Q^{\frac{1}{d-1}}}{R},
\end{align}
where $R$ is the radius of the sphere, \(\Lambda\) is the UV cut-off scale of the effective action, \(\rho\) is the charge density on a sphere, and \(Q\) is the total charge. This hierarchy allows the effective action in the large charge sector to be written perturbatively in $\frac{1}{Q}\ll 1$.
Assuming a homogeneous ground state with non-vanishing energy density in the large volume limit, $i.e.$ in the thermodynamic limit,
the effective action can be expressed in terms of the \(U(1)\) Goldstone mode, while all other massive excitations can be neglected, describing a superfluid state \cite{Hellerman:2015nra, Alvarez-Gaume:2016vff}. The ground state, determined by a \(U(1)\) charge \(Q\), has the ground state energy $E$ (or the corresponding scaling dimension $\Delta=ER$) as:
\begin{align}
    \Delta = c_1 Q^{\frac{d}{d-1}} + c_2 Q^{\frac{d-2}{d-1}} + \cdots.
\end{align}
The coefficients \(c_1\) and \(c_2\) are model-dependent parameters.

For a large angular momentum, on the other hand, a conformal bootstrap method has been used to determine the ground state energy. In particular, \cite{Alday:2007mf,Komargodski:2012ek, Fitzpatrick:2012yx} showed that the ground state energy in a CFT at large angular momentum is given by \begin{align} 
    \Delta = J + \Delta_{\Phi} Q + \cdots, 
\end{align} 
where $\Delta_{\Phi}$ is the dimension of the parton, an elementary constituent of the theory, with unit charge.
This large angular momentum regime, where the ground state can be described by this expression, is commonly referred to as the Regge limit. 
Physically, this can be understood as $Q$ partons rotating around the equator of the sphere with $J$ derivatives acting on the partons \cite{Alday:2007mf}. 
When the angular momentum is sufficiently large, interactions between the partons become negligible, making the total energy simply the sum of each parton’s energy.

A natural question arises: how does the ground state behave when both charge and angular momentum are large? Specifically, we are interested in the ground state’s behavior in the region between large charge and the Regge limit. This work will primarily examine the case in \( d = 3 \) (though the analysis can be generalized to higher dimensions, \( d > 3 \)).

With large angular momentum in addition to the large charge, the analysis becomes complex as spatial homogeneity is no longer expected. Additionally, for a superfluid state to host angular momentum, vortex configurations that carry singularities near the vortex cores must be taken into account.

In \cite{Cuomo:2017vzg}, the authors proposed the ground states for CFTs with \( U(1) \) charge on the sphere \( S^2 \times \mathbb{R} \) in the range \( Q^{1/2} \ll J \ll Q \) and $Q\ll J\ll Q^{3/2}$ by solving an EFT for vortex dynamics on \( S^2 \).
For \( Q^{1/2} \ll J \ll Q \), a single vortex–anti-vortex pair configuration emerges as the ground state, with the energy given by
\begin{align}\label{eq:1vort}
    \Delta=c_1 Q^{3/2}+\frac{\sqrt{Q}}{6c_1}\log \frac{J(J+1)}{Q}+\cdots.
\end{align}
For \( Q \ll J \ll Q^{3/2} \), many vortices appear, densely populated, and the superfluid’s flow resembles that of a (non-relativistic) rigidly rotating fluid:
\begin{align}\label{eq:rigid}
    \Delta=c_1 Q^{3/2}+\frac{1}{2c_1}\frac{J^2}{Q^{3/2}}+\cdots,
\end{align}
where corresponding the angular velocity is given as $\Omega R \approx \frac{J}{c_1Q^{3/2}} \ll 1$.
The effective field theory breaks down for $ Q^{3/2} \lesssim J$ as the angular velocity becomes relativistic $\Omega\lesssim 1/R$. 

This paper aims to deepen our understanding of universal charge relations within the charge regime $Q\ll J\ll Q^{2}$. Our main result is that the universal large charge expansion formula in this charge regime is given by
\begin{align}\label{eq:fluid}
    \Delta^2 = J^2 + c_1^2 Q^3 + \cdots
\end{align}
where $c_1$ is the same constant defined earlier.\footnote{As demonstrated in \S \ref{sub:validity}, the subleading terms arise at orders \( O(Q^2) \) and \( O(JQ) \). For \eqref{eq:fluid} to remain valid, both \( J^2 \) and \( Q^3 \) should be much larger than these subleading terms. This requirement translates to the condition \( Q \ll J \ll Q^2 \).}
Note that \eqref{eq:fluid} encompasses \eqref{eq:rigid}, as they coincide in the regime $Q \ll J \ll Q^{3/2}$.
Interestingly, we find that the energy expression in \eqref{eq:fluid} matches that of a normal fluid rigidly rotating with an angular velocity $\Omega$ in the zero-temperature limit. Not only does the functional form of the energy coincide, but the energy-momentum tensor and charge density also coincide. Despite the fundamental differences between superfluids and normal fluids—where the latter is characterized by macroscopic entropy while the former is described by a nearly unique ground state—this result suggests that the stationary configurations of the superfluid ground state can be effectively represented as a zero-temperature fluid. 

To be specific, our solution \eqref{eq:fluid} not only reproduces a slow rigid rotation phase with $\Omega\ll 1/R$ but describes a fast rigid rotation phase that corresponds to a rigid fluid rotating relativistically with $\Omega$ up to $1/R$.
In particular, the regime \( Q^{3/2} \ll J \ll Q^2 \) corresponds to a relativistic fluid rotating with angular velocity approaches as \( \Omega \to 1/R \). In this regime, the large charge expansion for the energy takes the form:  
\begin{align}\label{eq:fast rotating}  
    \Delta = J + \frac{c_1^2}{2} \frac{Q^{3}}{J}+\cdots.  
\end{align}  
We also investigate the excitation modes for \( Q^{3/2} \ll J \) and demonstrate the existence of chiral fluctuation modes, whose excitation energies scale as \( \delta E \sim \delta J/R \). As the angular momentum \( J \) increases for a fixed charge \( Q \), the system approaches the regime \( J \sim Q^2 \). In this limit, the chiral excitation modes with energy \( \delta E \approx \delta J/R \) are expected to condense, signaling a transition towards the Regge limit, where \( Q^2 \ll J \).

Apart from the correspondence between the superfluid ground state and the zero-temperature normal fluid established in this paper, the dynamics of conformal fluids are interesting by themselves due to the connection with the fluid-gravity correspondence (see, e.g., \cite{Policastro:2002se,Janik:2005zt,Son:2007vk,Bhattacharyya:2007vjd} and references therein).
For example, for holographic theories dual to Einstein-Maxwell gravity in AdS, zero-temperature fluid state corresponds to extremal black holes with event horizon radii \( r_+ \) much larger than the AdS radius. It has been known that large black holes in AdS\(_{d+1}\) can be effectively described by fluids on \( S^{d-1} \times \mathbb{R} \) \cite{Bhattacharyya:2007vs, Cardoso:2007ka}.  

In a different context, it was discovered that the functional form of the energy of extremal AdS$_4$-Reissner-Nordström black holes matches exactly with the ground-state energy of the linear sigma model with a \( U(1) \) global charge. This remarkable correspondence has been termed the AdS/EFT correspondence \cite{Loukas:2018zjh, Liu:2020uaz, delaFuente:2020yua}.
We suggest that the matching energy-charge relations (at least at leading order) arise from the fact that both extremal large AdS black holes and superfluid ground states can be described as zero-temperature fluids. This observation provides indirect evidence in favor of the AdS/EFT correspondence, even for systems with large angular momentum \( J \). 
However, it is important to highlight again the fundamental differences between superfluid and fluid descriptions, particularly regarding their entropies—where the former has zero entropy density, while the latter has a finite density—which suggest they are not entirely equivalent. For a detailed discussion of these distinctions, see \S \ref{sec: AdS EFT}.

As a final but important remark, we highlight that the authors of \cite{Cuomo:2022kio} also studied the large charge expansion in the regime \( Q^{3/2} \ll J \ll Q^2 \) and proposed that the ground state energy is given by  
\begin{align}\label{eq:giantvort}
    \Delta = J + \frac{9c_1^2}{4\pi} \frac{Q^3}{J} + \cdots.
\end{align}  
This configuration, referred to as the giant vortex state, describes a superfluid rotating around the sphere with angular velocity \(\Omega \to 1/R\), where the superfluid is supported only near the equator.  
We note that the fast rigid rotation phase shares qualitative similarities with the giant vortex state, particularly in the concentration of the stress-energy tensor and charge density around the equator. However, there is a notable difference in the coefficients appearing in \eqref{eq:fast rotating} and \eqref{eq:giantvort}, which results in the rigid rotation phase having lower energy compared to the giant vortex configuration.  
We will explore these similarities and differences in more detail in \S \ref{sub: giant vort}.

The rest of the paper is organized as follows:
Section \ref{sec: EFT review} provides a review of the large charge effective field theory on a sphere for the regime $J \ll Q^{3/2}$.  
Section \ref{sec: superfluid} presents the solution to the effective field theory for $Q \ll J \ll Q^2$ and establishes a connection with the non-relativistic limit. It also investigates fluctuations around the ground state and argues that these fluctuations condense as $J$ approaches $Q^2$, reaching the Regge limit.
Section \ref{sec: fluid} reviews conformal fluid dynamics, deriving the energy-charge relation for a zero-temperature fluid in terms of its charge and angular momentum. It also discusses holographic theories as concrete examples and study a relativistic fluid on a sphere with both large charge and angular momentum.   
Section \ref{sec: conc} concludes the paper and outlines directions for future research.

\section{Large Charge EFT with Spin}\label{sec: EFT review}
In this section, we briefly review the large charge EFT on the cylinder $\mathbb{R}\times S^2$ with metric $ds^2=-dt^2+R^2(d\theta^2+\sin^2\theta d\phi^2)$, at varying spins.

For a CFT with global $U(1)$ symmetry, an effective action for large charge ground state on the cylinder was written down by \cite{Hellerman:2015nra}, which is nothing but the action of a conformal superfluid:
\begin{align}\label{eq: effective action}
\mathcal{L} = \alpha |\partial \chi|^3 + \cdots.
\end{align}
Here $\chi\sim \chi+2\pi$ is the goldstone boson for the spontaneously broken $U(1)$ symmetry,
and \(|\partial \chi| = \sqrt{-\partial_{\mu} \chi \partial^{\mu} \chi}\).
The ellipsis represents higher derivative terms which are suppressed by $\frac{\partial^2}{|\partial\chi|^2}$.
The current and stress-energy tensor are:
\begin{align}\label{eq: superfluid current and stress tensor}
j_{\mu} = 3\alpha |\partial \chi| \partial_{\mu} \chi, \quad T_{\mu\nu} = \alpha \left(3 |\partial \chi| \partial_{\mu} \chi \partial_{\nu} \chi + g_{\mu\nu} |\partial \chi|^3 \right).
\end{align}
This effective action is supposed to be expanded around the static solution $\chi=-\mu t$, which is the large fixed charge ground state.
The $U(1)$ charge of this ground state is \(Q = 12\pi \alpha \mu^2R^2\), and the energy is \(E = 8\pi \alpha \mu^3R^2\).
Therefore, the relation between the ground state energy and charge can be written as:
\begin{align}
\Delta= c_{1} Q^{3/2} + c_{2} Q^{1/2} + \cdots,
\end{align}
where \(c_{1} = \frac{1}{3^{3/2} \sqrt{\pi \alpha}}\). The second term, with coefficient \(c_2\), arises from subleading terms in the action suppressed by $\frac{\partial^2}{|\partial\chi|^2}\sim\frac{1}{\mu^2R^2}\sim\frac{1}{Q}$.
Therefore, we observe that the higher-derivative terms are suppressed only when \( Q \) is large. In this regime, \(\mu \) serves as the effective field theory cutoff, enabling analytical analysis of the system.

Here we should note that we assume an `extensive' ground state, meaning that thermodynamic quantities like energy and charge scale proportionally with the volume of the sphere. 
(This is reflected in the EFT action \eqref{eq: effective action})
This assumption does not hold in theories with a moduli space of vacua, such as supersymmetric theories and free theories, where the ground state is non-extensive as was emphasized in e.g. \cite{Hellerman:2015nra,Hellerman:2017veg,Jafferis:2017zna,Cuomo:2024fuy}. 
For example, in supersymmetric theories with the moduli space,
the ground state for fixed large charge $Q$ is a BPS state satisfying
\begin{align}
    ER=\Delta = Q.
\end{align}
Extensivity cannot be maintained for a BPS state because if \(E \propto R^{d-1}\) and \(Q \propto R^{d-1}\), the left-hand side (\(R^d\)) and right-hand side (\(R^{d-1}\)) of the equation cannot match in the large $R$ limit. In fact the uniform energy density ($E/R^{d-1}$) and charge density ($Q/R^{d-1}$) must vanish in the large radius limit, and the analysis of \cite{Cuomo:2024fuy} implies that $Q\sim R^{d-2}$ and $E\sim R^{d-3}$.

To describe the theory away from the static phase, we expand the action around \(\chi = -\mu t\), such that \(\chi = -\mu t - \varphi\), where \(\varphi\) is the massless Goldstone mode. 
The effective action for the Goldstone mode around a static superfluid is given by
\begin{align}\label{eq: vortex eft}
    \sqrt{-g}\delta \mathcal{L}= 3\mu^2 \alpha \sqrt{-g}\partial_0\varphi+3\mu\alpha \sqrt{-g}\left[(\partial_0\varphi)^2-\frac{1}{2}(\partial_i\varphi)^2\right].
\end{align}
The first term is a total derivative, so it can typically be neglected. 

The fluctuations around the static superfluid follow the phonon dispersion relation characteristic of a conformal fluid in \(d=2+1\) dimensions, with a phonon velocity of \(\frac{1}{\sqrt{d-1}} = \frac{1}{\sqrt{2}}\), as observed from the quadratic terms in the effective action. The corresponding scaling dimension at large \(Q\) and finite \(J\) is then given by:
\begin{align}
\Delta = c_{1} Q^{3/2} + c_{2} Q^{1/2} + \frac{\sqrt{J(J+1)}}{\sqrt{2}} + \cdots.
\end{align}
This expansion is valid for angular momentum \(J \ll Q^{1/2}\). For higher angular momentum, the phonon description breaks down as the wavelength of the phonon approaches the cutoff length scale, leading to the proliferation of vortices, which we will explain shortly. Configurations incorporating vortices that solve the EFT \eqref{eq: vortex eft} for large angular momentum are discussed in \cite{Cuomo:2022kio}, and we briefly review them here.

First, we should emphasize that the vortex configuration is not single-valued and the phase around the core is quantized as $2\pi q$, where $q$ is the vorticity of the vortex.
To describe the configuration, we choose a normal coordinate $y^i(i=1,2)$ near the center of a vortex $y_0^i(t)$, such that $\varphi = \varphi(y^i-y^i_0(t))$. 
Near the vortex core, the linear term in \eqref{eq: vortex eft}, previously neglected for the phonon mode, becomes significant. This term can be expressed as:
\begin{align}
    3\mu^2 \alpha \int \sqrt{-g} d^3x \partial_0\varphi = 3\mu^2 \alpha \int dt d^2y \frac{\partial \varphi}{\partial y^i}\dot{y}^i_0(t).  
\end{align}
If the vorticity is $q$, the field $\varphi$ satisfies the following equation:
\begin{align}
    \partial_i\frac{\partial \varphi}{\partial y^j}- \partial_j\frac{\partial \varphi}{\partial y^i}=2\pi q \epsilon_{ij}\delta^2(y^i-y^i_0).
\end{align}
Therefore, the linear term can be written as
\begin{align}
    3\mu^2\alpha \int dt d^2y \frac{\partial \varphi}{\partial y^i}\dot{y}^i_0(t) =3\mu^2\alpha \int dt \pi q \epsilon_{ik}y_0^k \dot{y}^i_0(t).
\end{align}
This can be thought of as a world-line action for a particle with charge $q$:
\begin{align}\label{wrldact}
    q\int dt A_i(y(t))\dot{y}^i,
\end{align}
where the one-form field $A$ is defined to satisfy the following: $\partial_iA_j-\partial_iA_j=-6\alpha\mu^2\pi \epsilon_{ij}\sqrt{-g}$.
This picture manifests the vortex-particle duality. Namely, a vortex can be thought of as a particle on a uniform magnetic field and the vorticity corresponds to the electric charge of the particle.

On the other hand, the quadratic terms in \eqref{eq: vortex eft} can be thought of as the interactions between vortices.
For simplicity, we assume that the vortex velocity is sufficiently small (much less than the speed of light), allowing us to neglect the time derivative term. Under this assumption, the spatial derivative terms generate an electric potential between particles with charges.
The electric potential at a point particle located at \(r\) is:
\begin{align}\label{eq: electric potential}
    \phi(r) = -3\pi \mu \alpha\sum_{p}q_p\log\left((r - \vec{R}_p)^2\right),
\end{align}
where $r,\vec{R}_p$ are 3-vectors in the embedding space $\mathbb{R}^3$ of $S^2$.
$\vec{R}_p$ and $q_p$ are the location and charge of a $p-$th particle, respectively. 
The Lagrangian of the vortex can be expressed as:
\begin{align}\label{eq: effective superfluid}
     \delta \mathcal{L}= \sum_pq_p A_i(y_p) \dot{y}_p^i+3\pi\mu\alpha \sum_{p \neq r} q_p q_r \log\left(\mu|\vec{R}_p - \vec{R}_q|\right).
\end{align}
The energy and angular momentum that come from the vortex configuration are given by
\begin{align}
    E_{vort} &=-3\pi \mu \alpha\sum_{p\neq r}q_pq_r\log\left(\mu|\vec{R}_p - \vec{R}_r|\right),\nonumber\\
    \vec{J} &= \sum_{p} q_p \frac{Q}{2} \frac{\vec{R}_p}{R}.
\end{align}

Next, we examine stable configurations of vortices. The simplest scenario to consider is a vortex–anti-vortex pair. \footnote{Since the total vorticity on a closed surface must vanish, a single vortex cannot exist in isolation.}
For a vortex–anti-vortex configuration to be stable (stationary), the vortices must move on the sphere with a constant velocity. This motion generates a Lorentz force that counterbalances the electric force between the charged particles.
A static configuration can only occur when the vortex and anti-vortex are positioned at the north and south poles, respectively. If the distance between the two vortices is $L$, the energy and angular momentum of this configuration are given by:
\begin{align}
    E &= 8\pi R^2 \alpha \mu^3 + 6\pi \mu \alpha \log(\mu L), \nonumber\\
    J &= 6\pi R \mu^2 \alpha L.
\end{align}
where we have neglected the kinetic energy of the vortex since we are assuming a small vortex velocity.
The vortex-anti-vortex configuration (and its EFT) is valid for $L$ larger than the charge density scale $\frac{1}{\mu R}\sim \frac{1}{\sqrt{Q}}$, or equivalently, $Q^{1/2}\lesssim J$.
The large charge expansion of the vortex-anti-vortex configuration is written as
\begin{align}\label{eq: vav energy}
    \Delta=c_1 Q^{3/2}+\frac{\sqrt{Q}}{6c_1}\log\frac{J(J+1)}{Q}.
\end{align}
The maximum angular momentum for a vortex-anti-vortex pair is achieved in the static configuration, where each vortex is located at the poles, corresponding to $J=Q$.

To achieve higher angular momentum, two possible approaches emerge: (1) adding additional vorticity at each pole, or (2) increasing the total number of vortices.
Adding vorticity at each pole is energetically unfavorable because the self-energy of the vortex scales quadratically with the vorticity.
Therefore, the vortex-anti-vortex is valid for $Q^{1/2}\ll J \lesssim Q$ since the maximum angular momentum when vortex and anti-vortex are located at the pole scales with $J \sim O(Q)$.

As the angular momentum $J$ exceeds the charge $Q$, the number of vortices increases. Since increasing the vorticity of individual vortices is energetically unfavorable, the additional vortices all have a vorticity of $1$ or $-1$.
For $Q \ll J$, their distribution can be approximated as a continuous function of $\theta$ and $\phi$, as the vortices are densely distributed. The vortex distribution that minimizes the interaction energy \eqref{eq: electric potential} is given by \cite{Cuomo:2017vzg}:
\begin{align}\label{eq: rigid rotation vortex dist}
    \rho_v(\theta) = \frac{3}{2\pi R^2}\frac{J}{Q} \cos\theta,
\end{align}
where \(\rho_v(\theta)\) represents the vortex density. 
The equation of state is expressed as:
\begin{align}\label{nrlce}
    \Delta = c_1 Q^{3/2} + \frac{J^2}{2c_1 Q^{3/2}} + \cdots.
\end{align}
Interestingly, the energy of the ground state takes the same form as that of a rigidly rotating fluid with a constant angular velocity \( \Omega \ll 1 \) (see equation \eqref{eq: fluid rigid rot 2} for the equation of state of a zero-temperature fluid rotating with \( \Omega \ll 1 \)).  
In other words, the interaction energy between vortices—equivalent to the kinetic energy of a superfluid—shares the same form as the kinetic energy of a rigidly rotating fluid.

The effective field theory breaks down when the superfluid velocity reaches \( O(1) \) in units of the speed of light, corresponding to the angular velocity of the rigid fluid reaching \( O(1) \). This occurs when \( J \sim Q^{3/2} \).

\section{Relativistic Rigid Rotation Phase}\label{sec: superfluid}

For a larger angular momentum (\(Q^{3/2} \lesssim J\)), we must consider the relativistic effects that were previously neglected in \eqref{eq: effective superfluid}.
In this section, we show that the rigid rotation phase remains valid even within the relativistic limit.
Namely, we complete the rigid rotation phase to larger-$J$ regime. More specifically, we derive a new solution that is valid for $Q\ll J\ll Q^2$ which reduces to the previously described rigid rotation phase for $Q\ll J\ll Q^{3/2}$.
The energy, charge, and angular momentum of this solution satisfy:
\begin{align}\label{eq: superfluid lce}
    \Delta^2 = J^2 + c_1^2 Q^3+\cdots.
\end{align}
where $c_1=\frac{1}{3^{3/2}\sqrt{\pi\alpha}}$ is the same $c_1$ in the previous section.
\subsection{Emergent Degrees of Freedom}\label{new EFT}
In order to derive a new solution of
\begin{align}\label{eq: eom of EFT}
    \partial_{\mu}(\sqrt{-g}
    |\partial\chi|\partial^{\mu}\chi)=0
\end{align}
that minimizes the energy for given charge and angular momentum, we use the following ansatz: $\partial_\mu \chi(x) = \xi_\mu(x)$ where we treat $\xi$ as \emph{independent fundamental} variables.
This is because scanning over all possible vortex distributions give rise to \emph{emergent degrees of freedom} associated to the vortex density and its flow.
Let us elaborate on this point.

In the presence of vortices, $\chi$ is multi-valued by $2\pi$ times an integer amount: by going around a vortex with vorticity $q$, $\chi$ gets shifted by $2\pi q$. 
For the regime $Q \ll J \ll Q^{3/2}$, we have seen in the previous section that the ground state is a superfluid densely populated with vortices, allowing the vortex density to be approximated as a smooth function.
In this case, we essentially treat $\chi$ as multi-valued by $2\pi$ times any real number, because going around any circle will shift $\chi$ by $2\pi$ times the integral of vortex density inside that circle.
Consequently, $\chi$ itself lacks a well-defined value, and only $d\chi = \xi$ remains well-defined. 

Moreover, due to the smooth distribution of vortices, $\xi$ is no longer constrained by the condition $d\xi = 0$. Instead, $F \equiv d\xi$ serves as a measure of the vortex density and the flow of vortices in the following sense. In three dimensions, any closed 2-form $F$ gives rise to a trivially conserved current:
\begin{align}\label{eq: vorcurrent}
    j_v^\mu= \frac{1}{4\pi}\frac{\epsilon^{\mu\nu\rho}}{\sqrt{-g}}F_{\nu\rho}\;,
\end{align}
where $\epsilon^{012}=1$. This is covariantly conserved i.e. $\nabla_\mu j_v^\mu =\frac{1}{\sqrt{-g}}\partial_\mu(\sqrt{-g}j_v^\mu)=0$.
We call this vortex current, because its time component is nothing but the vortex density. Indeed, integrating the time component for some spatial region $A$,
\begin{align}
    \int_A d^2x\sqrt{-g}j^0_v
    =\frac{1}{2\pi}\int_A d^2x F_{12}
    =\frac{1}{2\pi}\oint_{\partial A}\xi_\mu dx^\mu,
\end{align}
we get the number of vortices (more precisely, total vorticity) inside the area $A$.
Since this is a conserved current, its spatial components should be interpreted as the flow of vortices. Note that integrating $j_v^0$ over the entire sphere yields zero, implying that the total vorticity must remain zero through any dynamical process.

This current, of course, also exists in the presence of \textit{finitely many} vortices.\footnote{Here, ``finitely many" means that the separation length scale between vortices is comparable to the length scale of interest, which is of order $O(R)$. In \S \ref{sub:validity}, we demonstrate that this condition is equivalent to $J \sim Q$.} In such cases, the components of the current are represented as sums of delta functions rather than smooth functions. Consequently, determining the ground state configuration with finitely many vortices amounts to fixing a finite set of parameters—such as the positions and vorticities of the vortices—along with the single-valued part of $\chi$. Since these finite parameters do not constitute a functional degree of freedom, the system has only one degree of freedom, which is associated with $\chi$, in the case of finitely many vortices.

When we can approximate the vortex density as smooth functions, on the other hand, we naturally approximate the whole vortex current as smooth functions. Therefore, finding a ground state configuration corresponds to fixing \emph{3 functions}: one is the single-valued part of $\chi$, and the remaining two are the components of the vortex current. Note that the vortex current is a conserved quantity hence out of 3 components, only 2 are independent degrees of freedom.
That is, we have 2 emergent degrees of freedom, which are components of the vortex current.

Consequently, $\xi_\mu$ can be treated as a new set of independent variables, detached from its origin as a gradient. This treatment, which we will refer to as \emph{effective description of vortices}, will later be shown in \S \ref{sub: perfect fluid} and \S \ref{sec: fluid} to be equivalent to a normal perfect fluid when combined with the continuity equations for $j^\mu$ and $T_{\mu\nu}$. 

Two points should be noted.  
First, the approximation of the vortex density as a smooth function neglects the mass of the vortices, which originates from the singularity in the field profile near the vortex core. As a result, this approximation is valid only when the total mass of the vortices is negligible at leading order. In \S \ref{sub:validity}, we will demonstrate that this condition holds for $Q \ll J \ll Q^2$. For the time being, we assume this validity and proceed with our analysis.

Second, since vortices are too heavy to be excited, the number of vortices remains fixed within the effective field theory. However, we note that there exist fluctuation modes of the vortex distribution itself within the EFT.\footnote{We thank Gabriel Cuomo for highlighting this point.} Consequently, when considering fluctuations around the ground state, it is necessary to account not only for the fluctuations of the single-valued component of $\chi$ but also for the fluctuations of the vortex current. This will be discussed in greater detail in \S \ref{sub: fluctuation}.

\subsection{The solution}\label{sub: E min}
The solution is determined as follows. 
We find $\xi_\mu(x)$ by minimizing the energy $E$ for fixed $J$ and $Q$. This procedure corresponds to determining the vortex distribution that minimizes the energy.
To achieve this, we use the Lagrange multiplier method i.e. we extremize the following quantity with respect to $\xi_\mu,\lambda_1,$ and $\lambda_2$.
\begin{align}\label{Lagrange multiplier}
    L\equiv E - \lambda_1 (J - J_0) - \lambda_2 (Q - Q_0).
\end{align}
A conserved charge associated with an isometry of spacetime is generally given by
\begin{align}
    Q[k^\mu] = \mathrm{sgn}(k^2)\int d^2x \sqrt{-g} T^{t\mu} g_{\mu\nu} k^{\nu},
\end{align}
where \( k^{\nu} \) is the corresponding Killing vector. Here, the sign factor is chosen to offset the negative sign of $g_{tt}$.
Hence the energy (associated to $\partial_t$) and angular momentum (associated to $\partial_\phi$) are
\begin{align}
    E &= R^2\int d\theta d\phi \sin\theta T_{tt}, \label{eq: E}\\
    J &= R^2\int d\theta d\phi \sin\theta (-T_{t\phi}),\label{eq: J}
\end{align}
and the charge is
\begin{align}
    Q &= R^2\int d\theta d\phi \sin\theta j^t,\label{eq: Q}
\end{align}
The stress-energy tensor and charge current are given by \eqref{eq: superfluid current and stress tensor}, which we reproduce here, with $\xi_\mu=\partial_\mu\chi$.
\begin{align}
    j^\mu=3\alpha |\xi|\xi^\mu\;,\quad
    T_{\mu\nu}=\alpha\big(
    3|\xi|\xi_\mu\xi_\nu+g_{\mu\nu}|\xi|^3\big)\;,
\end{align}
where $|\xi|=\sqrt{-\xi^\mu\xi_\mu}$.
Then, the Lagrange multiplier function $L$ is given by
\begin{align}
    L&=R^2\int d\theta d\phi \sin\theta\alpha f+\lambda_1 J_0+\lambda_2 Q_0\;,
\end{align}
where
\begin{align}
    f=&3|\xi|\xi_t^2-|\xi|^3
    -\lambda_1(-3|\xi|\xi_t\xi_\phi)
    -\lambda_2(-3|\xi|\xi_t)\;.
\end{align}
Using e.g. $\frac{\partial |\xi|}{\partial\xi_\mu}=-\frac{\xi^\mu}{|\xi|}$, it is straightforward to show that
\begin{equation}
\begin{aligned}
    \frac{\partial f}{\partial \xi_t}&=
    \frac{3}{|\xi|}(\xi_t^2+|\xi|^2)
    (\xi_t+\lambda_1\xi_\phi+\lambda_2)\;,\\
    \frac{\partial f}{\partial \xi_\theta}&=
    \frac{3\xi^\theta}{|\xi|}
    \big[|\xi|^2-\xi_t(\xi_t+\lambda_1\xi_\phi+\lambda_2)\big]\;,
    \\
    \frac{\partial f}{\partial \xi_\phi}&=
    \frac{3}{|\xi|}
    \Big[|\xi|^2(\xi^\phi+\lambda_1\xi_t)
    -\xi^\phi\xi_t(\xi_t+\lambda_1\xi_\phi+\lambda_2)\Big]\;,
\end{aligned}
\end{equation}
Setting these to zero, we get
\begin{equation}
\begin{aligned}
    \xi_t &= \frac{-\lambda_2}{1-\lambda_1^2R^2\sin^2\theta}\;,\\
    \xi_\theta &=0\;,\\
    \xi_\phi &= \frac{-\lambda_2}{1-\lambda_1^2R^2\sin^2\theta}
    (-\lambda_1 R^2\sin^2\theta)\;.
\end{aligned}
\end{equation}
If we decide to call $\lambda_1$ as $\Omega$ and $\lambda_2$ as $\mu$, we get the following solution:
\begin{align}\label{eq: the solution}
    \xi^\mu=\mu\gamma^2 (1,0,\Omega)\;,\;\;
    \gamma = \frac{1}{\sqrt{1-\Omega^2R^2\sin^2\theta}}\;.
\end{align}
This configuration (trivially) satisfies the equation of motion, $\partial_\mu(\sqrt{-g}|\xi|\xi^\mu) = 0.$ Additionally, the parameter $\Omega$ is constrained by an upper limit, $\Omega < 1/R.$ 
As we will discuss in \S \ref{sub: perfect fluid} and \S \ref{sec: fluid}, the normalized vector $u^\mu \equiv \xi^\mu/|\xi| = \gamma (1, 0, \Omega)$ can be interpreted as the 3-velocity of a relativistic normal fluid. In this interpretation, the angular velocity is given by $u^\phi/u^t = d\phi/dt = \Omega.$ Therefore, the velocity profile corresponds to a (relativistic) rigid rotation.

\begin{figure}
    \centering
    \includegraphics[width=1\linewidth]{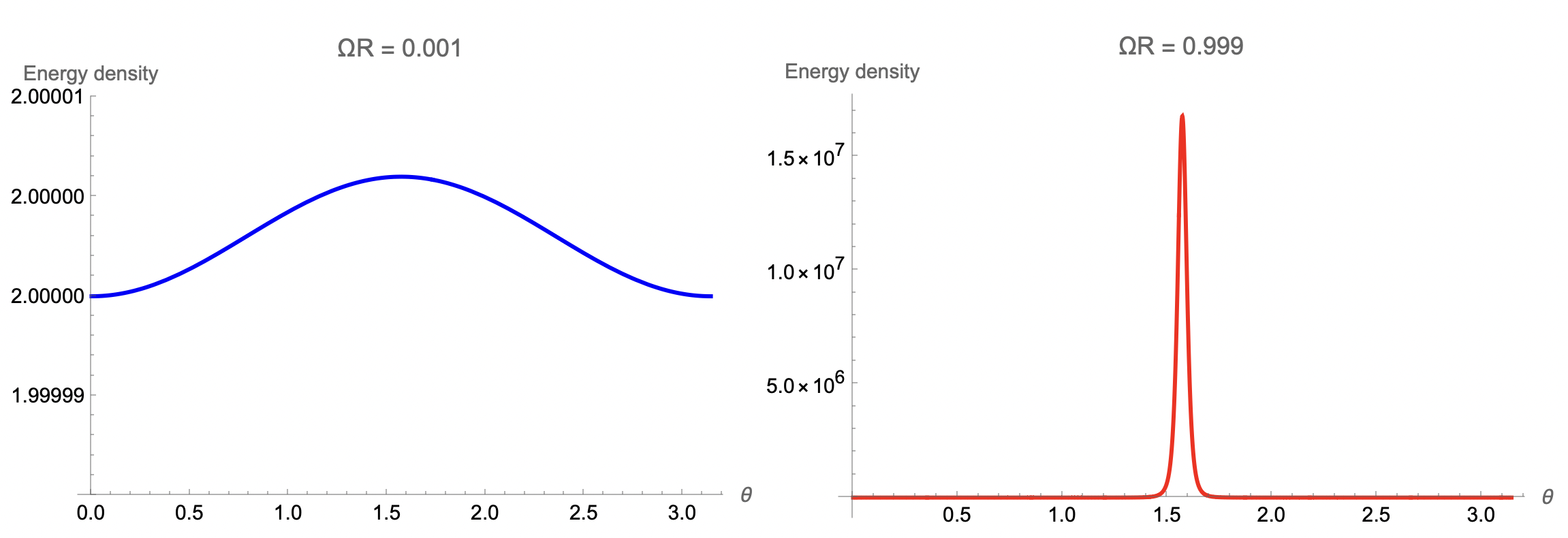}
    \caption{The energy density on a sphere as a function of $\theta$, the latitude angle, $0\leq\theta\leq \pi$ in units of $\alpha\mu^3$. Left: with $\Omega R=0.001$, where the energy density is almost uniform. Right: $\Omega R =0.999$. The energy density is concentrated on the equator.}
    \label{fig: energy den}
\end{figure}

Substituting $\xi$ into $J=J_0,\; Q=Q_0$ using \eqref{eq: J},\eqref{eq: Q}, one can determine $\Omega,\mu$ in terms of $J_0,Q_0$, which we will just call $J,Q$. 
The components of stress-energy tensor that is relevant to $E$ and $J$ are given as follows:
\begin{equation}\label{solstress}
\begin{aligned}
     T_{tt} = \alpha\mu^3\frac{2 + \Omega^2R^2 \sin^2 \theta}{(1-\Omega^2R^2\sin^2\theta)^{5/2}},~~~~~
     T_{t\phi} = -3\alpha \mu^3 \frac{\Omega R^2 \sin^2 \theta}{(1-\Omega^2R^2\sin^2\theta)^{5/2}}.
\end{aligned}
\end{equation}
Also, the charge density is given by
\begin{align}\label{solcharge}
    \rho=j^0=\frac{3\alpha\mu^2}{(1-\Omega^2R^2\sin^2\theta)^{3/2}}
\end{align}
Figure \ref{fig: energy den} displays the energy density on the sphere as a function of latitude angle \(\theta\). With \(\Omega=0.001\), the energy density is nearly uniform across the surface, while for \(\Omega=0.999\), it becomes highly concentrated near the equator, reflecting the effects of increased angular velocity.

Also, it is instructive to evaluate the vortex current \eqref{eq: vorcurrent} $j_v^\mu=\frac{\epsilon^{\mu\nu\rho}}{2\pi\sqrt{-g}}\partial_{\nu}\xi_\rho$ for the solution \eqref{eq: the solution}:
\begin{equation}
\begin{aligned}\label{eq: vordensity}
    \rho_v&\equiv j_v^t=\frac{\mu\Omega}{\pi }\frac{\cos\theta}{(1-\Omega^2R^2\sin^2\theta)^2}\;,
    \\
    j_v^\theta&=0\;,\\
    j_v^\phi&=\frac{\mu\Omega^2}{\pi }\frac{\cos\theta}{(1-\Omega^2R^2\sin^2\theta)^2}\;.
\end{aligned}
\end{equation}
The equation \eqref{eq: vordensity} agrees with the vortex distribution of the non-relativistic rigid rotation phase \eqref{eq: rigid rotation vortex dist} when $\Omega$ gets small (but not as small as $1/R^2\mu$, as we will see later in \eqref{eq: regime-lower}).
\begin{align}\label{eq: vorden2}
    \rho_v=\frac{3}{2\pi R^2}\frac{J}{Q}\cos\theta
    (1+O(\Omega^2R^2))\;.
\end{align}
Also, the equation \eqref{eq: vordensity} is consistent with the fact that the non-relativistic rigid rotation phase (which ignores flow of vortices) is valid for $J\ll Q^{3/2}$ because $j_v^\phi=\rho_v \Omega\sim \rho_v \frac{J}{RQ^{3/2}}$.
Note that the angular velocity of vortices (and anti-vortices on the southern hemisphere) is $j_v^\phi/j_v^t=\Omega$.

\begin{figure}
    \centering
    \includegraphics[width=1\linewidth]{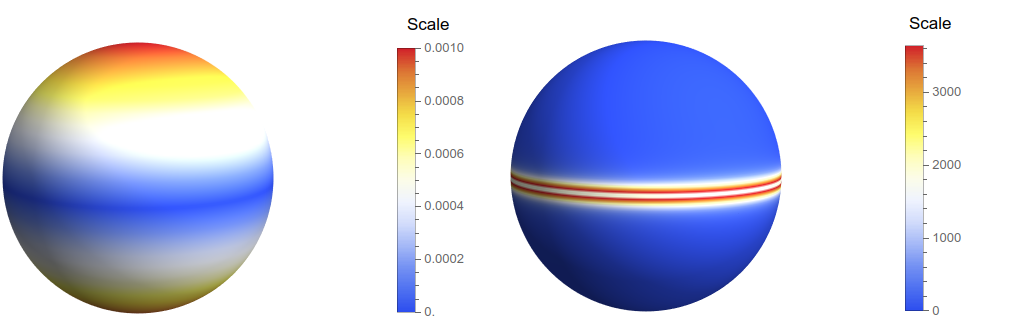}
    \caption{The absolute value of vortex density $R^2|\rho_v|$ on the sphere (we have set $\frac{\mu R}{\pi}=1$ for simplicity). Left: $\Omega R=0.001$. The vortex density is the largest at the poles, and vanishes at the equator.  Right: $\Omega R=0.999$. The vortex density is concentrated near the equator. The peaks are located at $\theta \approx \frac{\pi}{2} \pm \frac{\sqrt{1-\Omega^2R^2}}{\sqrt{3}}$.}
    \label{fig: vor dist}
\end{figure}

Figure \ref{fig: vor dist} shows the distribution of vortex density on a spherical surface, varying with angular velocity \(\Omega\). For small \(\Omega\), the vortex density peaks near the poles and decreases toward the equator. As \(\Omega\) approaches $1/R$, the vortex density is concentrated near the equator, with its peak located at $\theta \approx \frac{\pi}{2} \pm \frac{\sqrt{1-\Omega^2R^2}}{\sqrt{3}}$.

The energy, angular momentum, and charge of this configuration can be computed by plugging in \eqref{solstress}, \eqref{solcharge} into \eqref{eq: E}, \eqref{eq: J}, \eqref{eq: Q}. The result is
\begin{align}\label{eq: superfluid charges}
    E=\frac{8\pi\alpha R^2\mu^3}{(1-\Omega^2R^2)^2},\qquad 
    J=\frac{8\pi\alpha R^4\mu^3\Omega}{(1-\Omega^2R^2)^2},\qquad
    Q=\frac{12\pi\alpha R^2 \mu^2}{1-\Omega^2R^2}.
\end{align}
From this, we obtain an exact relation (up to vortex self energy contribution and higher derivative terms\footnote{
See \S \ref{sub:validity} for more details on the subleading corrections.}) among these quantities, given by the following equation:  
\begin{align}\label{superfluid lce2}
    \Delta^2= J^2 + c_1^2 Q^3
\end{align}
where $c_1=\frac{1}{3^{3/2}\sqrt{\pi\alpha}}$. This proves \eqref{eq: superfluid lce}.

\subsection{Non-relativistic limit and the worldline EFT}

It is straightforward to see that the non-relativistic limit of the energy-charge relation \eqref{superfluid lce2} reproduces the result obtained in \eqref{nrlce}. Similarly, the non-relativistic limit of the vortex density derived in \eqref{eq: vorden2} aligns with the result \eqref{eq: vordensity} in the regime where $\Omega \ll 1.$

In this subsection, we perform an additional consistency check by taking the new EFT introduced in \ref{new EFT}, where $\xi_{\mu}$ serves as the elementary degree of freedom, to the non-relativistic limit and comparing it with the vortex world-line EFT defined in \S \ref{sec: EFT review} at the Lagrangian level. This analysis confirms that the new EFT accurately captures all interaction energies between vortices and between a vortex and the magnetic field.

In the non-relativistic limit, the Lagrangian can be expanded perturbatively as:
\begin{align}
\mathcal{L}&=\alpha|\partial\chi|^3=\alpha\left((\mu+\dot{\varphi})^2-(\partial_i\varphi)^2\right)^{3/2}\\
    &=\alpha\mu^3+3 \alpha\mu^2 \partial_0\varphi+3\alpha\mu \left[(\partial_0\varphi)^2-\frac{1}{2}(\partial_i\varphi)^2\right]+\frac{\alpha}{2} \left(-3 \dot{\varphi} (\partial_i\varphi)^2+2 \dot{\varphi}^3\right)+\cdots.\label{nrl}
\end{align}
where $\chi=-\mu t - \varphi$. Since $\varphi$ is of order $O(\mu\Omega^2)$ the expansion can be thought of as an expansion with respect to $\Omega^2$, and thus is valid only in the non-relativistic limit where $\Omega R\ll 1$.
As explained in \S \ref{sec: EFT review}, the second term in equation \eqref{nrl}, $3\alpha\mu^2\partial_0\varphi$, corresponds to the world-line action $q\int dt A_i(y(t))\dot{y}^i,$ in \eqref{wrldact}. 

As a simple exercise, let us re-check this for the solution \eqref{eq: the solution}.
In the non-relativistic limit, the term $3\alpha\mu^2\partial_0\varphi$, on-shell, is expressed as:
\begin{align}\label{nron}
    3\alpha\mu^2\partial_0\varphi|_{\text{on-shell}}\approx -3\alpha\mu^2(\xi_0+\mu)\approx 3\alpha\mu^3\Omega^2R^2\sin^2\theta.
\end{align}
where we used $\xi_0=-\mu - \partial_{0}\varphi$. Integrating \eqref{nron} over the sphere yields:
\begin{align}
    \int_0^{\pi} 2\pi R\sin\theta d\theta 3\alpha\mu^3\Omega^2R^2\sin^2\theta = 8\pi\alpha\mu^3\Omega^2R^3.
\end{align}
On the other hand, the world-line action given in \eqref{wrldact} evaluates to:
\begin{align}
    \sum_{p}q_p\int A_{\mu}dX_p^{\mu}\approx \int 2\pi R^2\sin\theta d\theta  \rho_v \frac{Q}{2}(1-\cos\theta)\Omega = 8\pi\alpha\mu^3\Omega^2R^3,
\end{align}
which matches the above result. 

In the relativistic limit, higher-order terms in $\Omega$ in \eqref{nrl}, become relevant as $\Omega R$ approaches $O(1)$. 
From the perspective of the world-line EFT, writing down all such terms is theoretically possible but practically challenging. However, in the framework of our effective description of vortices, where the vortex density is treated as a fundamental degree of freedom, it becomes more straightforward to account for these terms—by writing down all possible terms with $\xi_\mu$. Due to the Weyl invariance, there is only one term at the leading order i.e. $|\xi|^3$. From this, all the coefficients of higher-derivative terms that are of order $\sim \mu^3$ are determined. Thus, all higher-derivative terms that are significant in the world-line EFT are systematically incorporated into our action.

It is worth noting, however, that our EFT assumes a smooth configuration of $\xi$. Consequently, the self-energy of the vortices must be accounted for separately, which we will address in the next subsection.

\subsection{Regime of Validity}\label{sub:validity}
Let us examine the regime of validity of this solution.
Firstly, we want the vortex density \eqref{eq: vordensity} to be much smaller than the charge density,
\begin{align}
    j^t=3\alpha|\xi|\xi^t=\frac{3\alpha\mu^2}
    {(1-\Omega^2R^2\sin^2\theta)^{3/2}}\;.
\end{align}
This is because the (local) length scale of the charge density corresponds to the UV cutoff of the original superfluid EFT, and we want the separation length scale between vortices—which serves as the new UV cutoff length scale—to be much larger than that.
\begin{align}
    \rho_v/j^t=\frac{1}{3\alpha\pi}\frac{\Omega}{\mu}\frac{\cos\theta}{(1-\Omega^2R^2\sin^2\theta)^{1/2}}
    \leq \frac{1}{3\alpha\pi}\frac{\Omega}{\mu}\;.
\end{align}
Last inequality can easily be seen by rewriting $\cos\theta$ as $(1-\sin^2\theta)^{1/2}$ and noting that $\Omega R<1$.
From this, we obtain the following simple condition,
\begin{align}
    \frac{\Omega}{\mu}\ll 1\;,
\end{align}
which is always satisfied because we require $\Omega<1/R\ll \mu$.
This is interesting because no matter how we take $\Omega$ close to $1/R$, which is the upper limit, vortices are not too close to each other as long as the chemical potential is large.
Also, we can rewrite this condition using equation \eqref{eq: superfluid charges} as
\begin{align}
    \frac{\Omega}{\mu}=18\pi\alpha\frac{J}{Q^2}\ll 1\;,
\end{align}
or, just simply as $J\ll Q^2$.

On the other hand, our effective description of superfluid vortices is reliable when the length scale of interest (which is $\sim R$) is much larger than the separation length scale between vortices which should be $\sim |\rho_v|^{-1/2}$. In other words, we require $1/R^2\ll |\rho_v|$.\footnote{
This might seem problematic at the equator, as $\rho_v$ vanishes. However, since the width of the strip near the equator where $|\rho_v|\sim 1/R^2$ is parametrically small (width $\sim \frac{(1-\Omega^2R^2)^2}{R\mu\Omega}\ll 1$), vortices are still typically separated by the length scale $\frac{(1-\Omega^2R^2)^2}{R\mu\Omega}$, because if we go out of this strip, there are large enough number of vortices.
} 
Taking a typical point on the sphere for the value of $\rho_v$, we get
\begin{align}\label{eq: regime-lower}
    \frac{1}{R^2}\ll \mu\Omega=\frac{3}{2R^2}(1-\Omega^2 R^2)\frac{J}{Q}\;,
\end{align}
Therefore, we see that we can't make $\Omega$ as small as $1/R^2\mu$ as mentioned before (above the equation \eqref{eq: vorden2}), and also, we see that our solution is valid for $Q\ll J$.

It is also important to note that the assumption of a smooth vortex current in the solution \eqref{eq: the solution}, representing a coarse-grained picture, ignores the self-energy (i.e. mass) of the vortices.  
In the fine-grained picture (in which we can distinguish each vortex), the field is singular at the core of a vortex, and this leads to the self-energy of a vortex. Thus, we must examine whether the self-energy of the vortices compromises the solution's validity as a candidate for the ground state configuration.
In the rest frame of the vortex, the self-energy is of order $O(\mu)$.  
For a rotating vortex with angular velocity $\Omega$, the self-energy becomes of order $O(\gamma \mu)$, where $\gamma = \frac{1}{\sqrt{1-\Omega^2 R^2}}$.  
Therefore, the total self-energy and self-angular momentum of the vortices $(E_v, J_v)$ are written as:  
\begin{equation}
\begin{aligned}
    E_{v}&=c_{\mathrm{self}}\int d^2x\sqrt{-g} \gamma \mu ~ |j_{v}^{t}|=\frac{2c_{\mathrm{self}}\Omega(3-2\Omega^2 R^2)\mu^2R^2}{3(1-\Omega^2 R^2)^{3/2}},\\
    J_{v}&=c_{\mathrm{self}}\int d^2x\sqrt{-g} \gamma \mu ~ |j_{v}^{\phi}|~ R^2\sin^2\theta=\frac{2c_{\mathrm{self}}\Omega^2 R^4\mu^2}{3(1-\Omega^2R^2)^{3/2}}.
\end{aligned}
\end{equation}
where $c_{\mathrm{self}}$ is some order $1$ theory-dependent constant (a new Wilson coefficient) and $j_v^{\mu}$ is given in \eqref{eq: vordensity}.
We computed $J_v$ from $J_v=\sum_i m_i \vec{r}_i\times \vec{v}_i $.
One can observe that the self-energy and angular momentum are much smaller than the leading-order energy and angular momentum written in \eqref{eq: superfluid charges}.  More specifically,
\begin{equation}
\begin{aligned}
    \frac{E_v}{E}\sim\frac{J_v}{J}\sim \frac{\Omega (1-\Omega^2R^2)}{\mu}.
\end{aligned}
\end{equation}
Therefore, accounting for the self-energy of the vortices does not require changing the solution at the leading order.  
Moreover, it can be shown that the energy-charge relation \eqref{eq: superfluid lce}, $\Delta^2-J^2-c_1^2Q^3=0$, remains unchanged.
The change to the equation due to the self-energy contribution from the vortices, is:  
\begin{align}
    \delta (\Delta^2 - J^2) \approx 2\Delta E_v R - 2J J_v = O\left(\frac{\Omega R^6 \mu^5}{(1-\Omega^2R^2)^{5/2}}\right),
\end{align}  
which is always much smaller than $c_1^2 Q^3$ that is of order $O\left(\frac{\mu^6 R^6}{(1-\Omega^2R^2)^{3}}\right)$.
Therefore at the leading order, one can still trust the equation \eqref{eq: superfluid lce}.

Finally, we compute how much higher derivative terms—which were neglected in the effective action \eqref{eq: effective action}—are suppressed.
The exact forms of those terms can be found in e.g. \cite{Hellerman:2015nra},\cite{Monin:2016jmo},\cite{Cuomo:2022kio}. 
Let us use the expression of \cite{Cuomo:2022kio} and evaluate each term for the solution \eqref{eq: the solution}.
There are 5 terms at the $O(\partial^2/|\partial\chi|^2)$ order relative to the leading term $|\partial\chi|^3$.
Each of them evaluates to:
\begin{equation}
\begin{aligned}
    \frac{\mathcal{R}}{|\partial\chi|^2}
    =&\frac{2}{R^2\mu^2}(1-\Omega^2R^2\sin^2\theta)
    ,\\
    \frac{\nabla^2(|\partial\chi|^{1/2})}{|\partial\chi|^{5/2}}
    =&\frac{1}{R^2\mu^2}\bigg[-\frac{\Omega^2R^2}{4}\sin^2\theta
    +\frac{1}{4}(5-\Omega^2R^2)
    -\frac{5}{4}\frac{1-\Omega^2R^2}{1-\Omega^2R^2\sin^2\theta}
    \bigg]
    ,\\
    \mathcal{R}_{\mu\nu}
    \frac{\partial^\mu\chi\partial^\nu\chi}{|\partial\chi|^4}
    =&\frac{\Omega^2R^2\sin^2\theta}{R^2\mu^2}
    ,\\
    \frac{\nabla^2|\partial\chi|}{|\partial\chi|^3}
    =&\frac{1}{R^2\mu^2}\bigg[3-\Omega^2R^2
    -\frac{3(1-\Omega^2R^2)}{1-\Omega^2R^2\sin^2\theta}
    \bigg]
    ,\\
    \frac{\partial^\mu\chi\partial^\nu\chi
    \nabla_\mu\nabla_\nu(|\partial\chi|^{-1})}{|\partial\chi|^3}
    =&\frac{1}{R^2\mu^2}\bigg[
    -\Omega^2R^2\sin^2\theta
    -(1-\Omega^2R^2)
    +\frac{(1-\Omega^2R^2)}{1-\Omega^2R^2\sin^2\theta}
    \bigg].
\end{aligned}
\end{equation}
From this, we see that all the terms are of order $\frac{1}{R^2\mu^2}\ll 1$ compared to the leading term.
Here, we have written down the right hand sides such that every term manifestly does not diverge near the equator as $\Omega\to 1/R$.
Physically, the suppression of these terms as $\Omega \to 1/R$ arises from the fact that most vortices are concentrated near the equator. Higher-derivative terms of this kind are associated with vortex acceleration, but since the vortices' motion is nearly geodesic, these terms are suppressed.\footnote{We thank Gabriel Cuomo for offering this microscopic interpretation.}

In summary, in order for our solution \eqref{eq: the solution} to be valid, we require $1/R\ll \sqrt{|\rho_v|}\ll \sqrt{\rho}$ which translates to $Q\ll J\ll Q^2$.

\subsection{Perfect fluid}\label{sub: perfect fluid}

We started from the EFT of conformal superfluid \eqref{eq: effective action} and then proposed a new description of vortices (plus the single-valued part of $\chi$) in terms of fundamental variables $\xi_\mu$.
In fact, this effective description of vortices can easily be shown to be equivalent to a (normal) perfect fluid, when our system is seen at late times and large distances.



Recall that our charge current and stress tensor are given by
\begin{align}
    j^\mu=3\alpha |\xi|\xi^\mu\;,\quad
    T_{\mu\nu}=\alpha\big(
    3|\xi|\xi_\mu\xi_\nu+g_{\mu\nu}|\xi|^3\big)
    \;,
\end{align}
On the other hand, perfect fluid's charge current and stress tensor are given as \eqref{eq: perfect fluid constituitive}:
\begin{align}
(j^{\mu})_{\mathrm{perfect}}=\rho u^\mu,\quad
(T^{\mu\nu})_{\mathrm{perfect}}=\epsilon u^\mu u^\nu +\mathcal{P} P^{\mu\nu}
\end{align}
where $u^\mu$ is the normalized 3-velocity of the superfluid, $\rho$ is the proper charge density\footnote{Note that this charge density $\rho=\pm|j^\mu j_\mu|^{1/2}$ is measured from a local observer flowing with the fluid, which is in general different from just the time component of the current, a coordinate dependent quantity.}, $\mathcal{P}$ is the pressure, and $\epsilon$ is the proper energy density. From the traceless condition, we have $\mathcal{P}=\epsilon/2$. We also defined  $P^{\mu\nu}=g^{\mu\nu}+u^{\mu}u^{\nu}$. These kinds of details are further explained in \S \ref{sec: fluid}.

One can immediately see that the current and stress tensor are exactly in the form of perfect fluid with the identification
\begin{equation}
\begin{aligned}
    u^\mu = \frac{\xi^\mu}{{|\xi|}},\\
    \rho = 3\alpha|\xi|^2,\\
    \epsilon = 2\alpha |\xi|^3.
\end{aligned}
\end{equation}
Also, the dynamics of a perfect fluid is determined by the continuity equations of the charge current and the stress tensor. These equations hold in our case due to the EOM obtained from the variation of the single-valued part of $\chi$.

Obviously, this identification does not care about whether the curl-free condition $d\xi=0$ is violated or not. 
This is just the reflection of the well-known fact that our starting point (EFT of $U(1)$ goldstone boson) is the theory of zero-temperature superfluids \cite{Schmitt:2014eka}, whose equation of state is fixed by conformal invariance.

\subsection{Fluctuations and Phase transition to the Regge limit}\label{sub: fluctuation}
Here, we analyze the fluctuations around the fast rigid rotation phase ($\Omega \to \frac{1}{R}$) by perturbing the field around the solution \eqref{eq: the solution}. 
We propose that a fluctuation mode, which is stable in the rigid rotation phase, become unstable as $J$ approaches $Q^2$, signaling a transition to the Regge limit.

We introduce the fluctuation field $\varphi$ with $\chi \to \chi + \varphi$, where the term $|\partial\chi|^3$ transforms as follows:
\begin{align}
    |\partial\chi|^3\to
    |\partial\chi|^3-3|\partial\chi|\partial^\mu\chi\partial_\mu\varphi
    -\frac{3}{2}|\partial\chi|\partial_\mu\varphi\partial^\mu\varphi
    +\frac{3}{2|\partial\chi|}\partial^\mu\chi\partial^\nu\chi\partial_\mu\varphi\partial_\nu\varphi+O(\varphi^3),
\end{align}
Plugging in $d\chi=\mu\gamma^2(-dt+\Omega R^2\sin^2\theta d\phi)$, ($|\partial\chi|=\mu\gamma$, $\partial^\mu\chi\partial_\mu\varphi=\mu\gamma^2(\partial_t\varphi+\Omega\partial_\phi\varphi)$) we get the Lagrangian as follows
\begin{align}
    \mathcal{L}=\alpha\mu^3\gamma^3
    -\frac{3\alpha\mu}{2}\gamma \partial_\mu\varphi\partial^\mu\varphi
    +\frac{3\alpha\mu}{2}\gamma^3(\partial_t\varphi+\Omega\partial_\phi\varphi)^2+O(\varphi^3),
\end{align}
Here, we are going to consider single-valued fluctuations only, hence we dropped the linear term which becomes a total derivative due to EOM. In other words, we do not consider fluctuations in vortex current.
\footnote{We note that there might be additional light modes arising from fluctuations in the vortex densities, which cannot be captured by single-valued fluctuations \cite{Cuomo:2021qws}. Since the mass scales of the vortex lattice fluctuation and the single-valued mode differ at leading order, we assume that the mixing with vortex lattice fluctuations is small enough to be neglected at this order. However, this assumption needs to be justified through concrete computation, which we leave as a future study. We thank G. Cuomo for pointing out this possibility.}

We redefine $\theta$ variable as $\cos\theta\equiv\sqrt{(1-\Omega^2R^2)}x$ and further redefine the variable as
\begin{align}
    \frac{x}{(1+x^2)^{1/2}}=y, \frac{dx}{(1+x^2)^{3/2}}=dy, 
\end{align}
where the domain of $y$ is $\left[-\frac{1}{\sqrt{2-\Omega^2R^2}},\frac{1}{\sqrt{2-\Omega^2R^2}}\right]$.
The effective action is written as
\begin{align}
    \frac{3\alpha\mu}{2(1-\Omega^2R^2)^{3/2}}\int dt dy d\phi\left[(\partial_0\varphi+\Omega\partial_{\phi}\varphi)^2-(1-y^2)^2 \partial_{y}\varphi\partial_{y}\varphi-\frac{1-\Omega^2R^2}{1-y^2}(-\partial_{0}\varphi\partial_{0}\varphi+\partial_{\phi}\varphi\partial_{\phi}\varphi)\right]
\end{align}
The equation of motion $\frac{\partial \mathcal{L}}{\partial \varphi}=\partial_{\mu}\frac{\partial \mathcal{L}}{\partial(\partial_{\mu}\varphi)}$ is written as:
\begin{equation}
\begin{aligned}\label{eq: eom of fluc}
    (\partial_0+\Omega\partial_{\phi})^2\varphi-\partial_y\left[(1-y^2)^2\partial_y\varphi\right]-\frac{1-\Omega^2R^2}{1-y^2}(-\partial_{0}^2\varphi+\partial_{\phi}^2\varphi)=0.
\end{aligned}
\end{equation}

At the leading order in $O\left(1-\Omega R\right)$, and assuming a solution independent of $y$, the EOM simplifies to
\begin{equation}
\begin{aligned}\label{eq: leading eom of fluc}
& (\partial_0 + \frac{1}{R} \partial_{\phi})^2 \varphi = 0,  \\
& J = m, \quad E \approx \frac{m}{R}.
\end{aligned}
\end{equation}
where $m$ is an integer.
This describes chiral (de)excitation modes, which can be interpreted as modes rotating on the sphere at the speed of light. Since there are infinitely many such chiral modes at leading order, they can combine to form an infinite vacuum degeneracy. However, this degeneracy is expected to be lifted when subleading corrections are taken into account \cite{Cuomo:2022kio}.\footnote{The mode dependent on $y$ is massive, whose energy $E$ is strictly larger than the absolute value of the angular momentum $|J/R|$. Therefore it does not contribute to the degeneracy of the vacuum.}

The rigid rotation phase should remain stable under fluctuations of the form $\chi \rightarrow \chi + \varphi$, as the energy of the vortex configuration has been explicitly extremized (see \S \ref{sub: E min}). If this were not the case, the chiral excitation mode would condense, causing a breakdown of the superfluid phase—a scenario we do not expect.
Thus, the energy cost of exciting the chiral mode must exceed the energy required to remain in the superfluid phase at the same angular momentum. Similarly, the energy reduction from de-excitation must be smaller than the corresponding energy decrease in the superfluid phase.

To investigate this, we consider the energy of the large charge expansion for the rigid rotation phase with \(\Omega \rightarrow \frac{1}{R}\):
\begin{align}
    \Delta = J + \frac{c_1^2}{2} \frac{Q^3}{J}.
\end{align}
As \(J\) increases by \(\delta J\), the change in energy \(\delta E\) with fixed charge $Q$ is given by:
\begin{align}
    \frac{\delta E }{\delta J} = \frac{1}{R}\left(1 - \frac{c_1^2}{2} \frac{Q^3}{J^2}\right).
\end{align}
Since we are assuming $Q^{3/2}\ll J$, the leading-order energy-to-angular momentum ratio is $1/R$, matching that of the chiral mode. This indicates that the superfluid solution is marginally stable at the leading order.

To confirm the stability, one has to study the subleading corrections to the chiral fluctuations from higher order derivative terms and vortex density fluctuations to prove  
\begin{align}
    \frac{\delta E_{-}}{\delta J} < \frac{1}{R}\left(1 - \frac{c_1^2}{2} \frac{Q^3}{J^2}\right) < \frac{\delta E_{+}}{\delta J},
\end{align}
where $E_{\pm}$ are the lightest excitation and de-excitation energies, respectively.
However, to find the full spectrum of the fluctuation turns out to be very difficult.
As a consistency check, we note the presence of a conformal descendant mode with $\frac{\delta E}{\delta J}=\frac{1}{R}$. Since this mode only increases the energy (so that it offers an upper bound of $\frac{\delta E_+}{\delta J}$), it does not contradict our identification of the ground state.

What happens between the rigid rotation phase $J\ll Q^2$ and the Regge limit $Q^2\ll J$ is an interesting question. Here, we address it in a rather qualitative way. 
Consider fixing \( Q \) and increasing \( J \) by raising the angular velocity \( \Omega \rightarrow 1/R \). Since the charge \( Q \) is fixed, the chemical potential \( \mu \) must decrease.
At some point, \( \mu \) is no longer large and becomes of order \( 1 \). As a result, the EFT breaks down, and the large charge expansion of the energy, as described by equation \eqref{eq: superfluid lce}, is no longer valid.
In the regime where $J \sim Q^2$, chiral fluctuation modes are expected to condense as $\Omega \to 1/R$. For small $\mu\lesssim 1$,  we expect $\frac{\delta E}{\delta J}$ to exceed $1/R$. 
At this point, it becomes energetically favorable to excite chiral modes with $E = J$ to approach the Regge limit, where the phase is governed by chiral modes of partons. 

Interestingly, a similar phenomenon occurs for black holes in AdS. Consider a black hole with \( \Omega R \approx 1 \) and \( \mu R < 1 \). At fixed charges, the slope \( \frac{\partial E}{\partial J} \) of the black hole is always greater than \( 1/R \), making it energetically favorable for chiral modes to appear.  
In AdS, the condensation of the chiral modes can be interpreted as a form of ``hair" that rotates around the black hole with angular velocity \( \Omega \to 1/R \). These modes are located far from the black hole's center, allowing the hair and the core of the black hole to interact minimally.  
This type of gravitational configuration has recently been constructed and is referred to as the ``Grey Galaxy" \cite{Kim:2023sig}.

Since the superfluid state does not have a macroscopic number of states, it is possible that the superfluid does not correspond directly to a geometry that has an event horizon. However, even if this is not the case, the large-charge superfluid state might be understood as a heavy bosonic star \cite{Liu:2020uaz}. We believe that the grey galaxy analogy provides a framework for understanding our setup as well.

\subsection{Remarks on giant vortex solution}\label{sub: giant vort}

We point out an additional solution to the EFT \eqref{eq: effective action}, referred to as the ``giant vortex," which arises in the regime $Q^{3/2} \ll J \ll Q^2$ \cite{Cuomo:2022kio}.
The giant vortex solution is described by the ansatz
\begin{align}\label{eq: gvor ansatz}
     \chi = -\mu t - l \phi.
\end{align}
The ansatz remains valid on the sphere as long as $-\partial_{\mu} \chi \partial^{\mu} \chi$ is positive.
The domain of validity is therefore confined to the region where $-\partial_{\mu} \chi \partial^{\mu} \chi$ is positive, which corresponds to $\frac{\pi}{2} - \delta < \theta < \frac{\pi}{2} + \delta$, where $\cos\delta = \frac{|l|}{R\mu}$. Outside this domain, the field must vanish. The boundary condition $j_\theta=0$ is imposed on the boundary, which is the only consistent condition that can be imposed \cite{Cuomo:2021cnb}.

From \eqref{eq: superfluid current and stress tensor} and \eqref{eq: gvor ansatz}, it is straightforward to obtain stress-energy tensor and current density, which is given as
\begin{equation}
\begin{aligned}
    j^t&=3\alpha\mu^2 R^2\sqrt{1-\frac{\cos^2\delta}{\sin^2\theta}},\\
    T_{tt}&=\alpha\mu^3 R^2 \left[3\sqrt{1-\frac{\cos^2\delta}{\sin^2\theta}}-\left(\frac{\cos^2\delta}{\sin^2\theta}\right)^{3/2}\right].
\end{aligned}
\end{equation}
for $\frac{\pi}{2} - \delta < \theta < \frac{\pi}{2} + \delta$. Upon integrating the densities, we obtain the relation between $E, J$, and $Q$ (Note that $J$ is given as $J=lQ$).
\begin{align}
    \Delta=J+\frac{9c_1^2}{4\pi}\frac{Q^3}{J}+\cdots
\end{align}
Given that the coefficient is larger $\frac{9c_1^2}{4\pi}$ than that of the fast-rotating superfluid $\frac{c_1^2}{2}$, we believe that the giant vortex cannot be the ground state.

Now, let us compare the energy density profile of the giant vortex and rigid rotation phases. For the giant vortex, the energy density can be rewritten as
\begin{equation}
\begin{aligned}
    T^{tt} &\propto \sqrt{1 - \frac{\cos^2 \delta}{\sin^2 \theta}} \left(2 + \frac{\cos^2 \delta}{\sin^2 \theta}\right), \quad \text{for} \quad \frac{\pi}{2} - \delta < \theta < \frac{\pi}{2} + \delta, \\
    T^{tt} &= 0, \quad \text{elsewhere},
\end{aligned}
\end{equation}
where \( \delta = \frac{1}{\pi \sqrt{3\alpha}} \frac{Q^{3/2}}{J} \).
For the relativistic rigid rotation phase, on the other hand, as the angular velocity \( \Omega \) approaches 1, which is equivalent to \( Q^{3/2} \ll J \ll Q^2 \), the $\theta$ dependence on the energy density is given as
\begin{align}
    T^{tt} \propto \frac{1}{(1 - \Omega^2 + \Omega^2 \cos^2 \theta)^{5/2}} \sim \frac{1}{\left(\frac{1}{27 \alpha \pi} \frac{Q^3}{J^2} + \cos^2 \theta\right)^{5/2}}.
\end{align}
Near the equator, the energy density decays as quickly as
\begin{align}
    T^{tt} \sim \frac{1}{\left(\frac{\delta^2}{9}+\left(\frac{\pi}{2}-\theta\right)^2\right)^{5/2}}
\end{align}
Both solutions are similar in that the energy density is concentrated around the equator, with the width of the strip being of order \( \delta \sim Q^{3/2}/J \).
However, a key difference is that in our solution, the energy density is smooth at finite \( \delta \), whereas in the giant vortex solution, the energy density is non-smooth.

It would be interesting to verify the profile of the ground state energy density by experiment or numerical analysis. 
There have been some numerics \cite{Kasamatsu_2002,Fetter_2005} and experiments \cite{Madison:2000zz,bretin2003fast,aftalion2004giant} 
on the vortex configuration, by considering a bose einstein condensation trapped in a confining potential with sufficiently large angular momentum. They both find that at the fast rotating regime, the vortex configuration looks very similar to the fast rotating/giant vortex configuration.
However, since \( \delta \) is a very small parameter for \( \Omega \approx 1/R \) (with \( \delta^2 \propto 1 - \Omega R \)), high resolution is required to capture the details accurately.

\section{Stationary normal fluid}\label{sec: fluid}

As discussed in \S\ref{sub: perfect fluid}, the emergent three degrees of freedom can be interpreted as those describing a perfect fluid. In this section, we demonstrate that the exact solution \eqref{eq: the solution} can also be derived directly from fluid dynamics. Specifically, we reproduce the solution \eqref{eq: the solution} by taking the zero-temperature limit of a general stationary solution of a fluid on \(\mathbb{R} \times S^{d-1}\), as obtained in \cite{Bhattacharyya:2007vs}.  

We further show that the zero-temperature limit of the stationary fluid solution can be connected to extremal black hole solutions, as large AdS black holes can be effectively described as conformal fluids via the fluid-gravity correspondence. As a concrete example, we demonstrate that the energy and charge structure of large extremal AdS$_4$ Kerr-Newman black holes aligns with that of fluid dynamics on a sphere with angular momentum.  

Finally, we compare the energy and charge structures of the superfluid ground state, the zero-temperature fluid, and the extremal black hole, and discuss their implications for the AdS/EFT correspondence.

\subsection{Review of fluid dynamics}
In this subsection, we review the relativistic (conformal) fluid mechanics, following \cite{Bhattacharyya:2007vs}. 
Let us first consider a QFT on $d$-dimensional spacetime, with several conserved currents.
While our primary focus will be on CFTs defined on $\mathbb{R} \times S^{d-1}$ with radius $R$, for the purpose of this subsection, we assume a general manifold and a generic QFT.

Let us consider generic states with finite energy density supported across the entire system and suppose we focus on physics at late times and over long distances.  
In this regime, it is reasonable to assume that the dynamics are governed entirely by conserved currents, as these are protected by continuity equations. All other microscopic excitations are short-lived and eventually decay into functions of the conserved currents and their derivatives.  
Consequently, to describe the state of a QFT with finite energy density at late times, it suffices to consider the stress tensor \( T^{\mu\nu} \) and the charge currents \( J_i^\mu \), where \( i = 1, 2, \cdots \) label the conserved currents.

Fluid mechanics goes a step further.  
It describes systems in \emph{local} thermodynamic equilibrium.  
Thus, instead of the full stress tensor and charge currents, it treats the local proper energy density \(\epsilon\) (\(=T^{00}\)), local charge densities \(\rho_i\) (\(=J_i^0\)), and fluid velocities \(u^\mu = \gamma(1, \vec{v})\), where \(u_\mu u^\mu = -1\), as the fundamental degrees of freedom. The stress tensor and charge currents are then expressed as functions of \(\epsilon\), \(\rho_i\), \(u^\mu\), and their derivatives. Such functional relations are referred to as \emph{constitutive relations}. 
These local quantities \(\epsilon\), \(\rho_i\), and \(u^\mu\) completely specify a fluid dynamical state, and the equations of motion are given by the continuity equations for the stress tensor and charge currents.  
Due to local thermodynamic equilibrium, it is often convenient to use the local temperature \(\mathcal{T}\) and local chemical potentials \(\mu_i\) as variables instead of energy density and charge densities, and we will adopt this convention frequently.

\subsubsection{Constitutive relations}
As an effective theory valid at late times and long distances, fluid mechanics admits a derivative expansion. This expansion can also be understood as an expansion in the mean free path \(l_{\mathrm{mfp}}\), which serves as the UV cutoff of the effective description.  
The expansion parameter is determined by the ratio \(l_{\mathrm{mfp}}/L_{\mathrm{fluid}}\), where \(L_{\mathrm{fluid}}\) represents the typical length scale of the fluid solution and sets the scale of the derivatives.  
As we will see later, the fluid length scale \(L_{\mathrm{fluid}}\) corresponds to the radius of the sphere \(R\) in our case.

To zeroth order in derivative, Lorentz invariance and the correct static limit uniquely fix $T^{\mu\nu}$, $J_i^{\mu}$, and the entropy current $J_{S}^{\mu}$ as functions of fundamental fluid mechanics variables.
In other words, constitutive relations are uniquely fixed at zeroth order as follows.
\begin{equation}
\begin{aligned}\label{eq: perfect fluid constituitive}
&T^{\mu\nu}_{\mathrm{perfect}}=\epsilon u^\mu u^\nu +\mathcal{P} P^{\mu\nu},\\
&(J_i^{\mu})_{\mathrm{perfect}}=\rho_i u^\mu,\\
&(J_S^{\mu})_{\mathrm{perfect}}=s u^\mu,
\end{aligned}
\end{equation}
where $s(\mathcal{T},\mu_i)$ is the rest frame entropy density.
Here, we defined a projector,
\begin{align}
    P^{\mu\nu}=g^{\mu\nu}+u^\mu u^\nu\;,
\end{align}
which projects vectors onto a subspace orthogonal to $u^\mu$.
Also, the coefficient of this projector, $\mathcal{P}$, is nothing but the pressure.
At this order, there is no dissipation, no entropy production.
Therefore, the fluids described by \eqref{eq: perfect fluid constituitive} are called perfect fluid.

At the first-order in derivative, constitutive relations are given as (see e.g. subsection 14.1 of \cite{Andersson:2006nr} for derivations)
\begin{equation}
\begin{aligned}\label{eq: first-order constitutive}
    &T^{\mu\nu}_{\mathrm{dissipative}}=-\zeta\vartheta P^{\mu\nu}-2\eta \sigma^{\mu\nu}+q^{\mu}u^{\nu}+u^{\mu}q^{\nu},\\
    &(J_i^{\mu})_{\mathrm{dissipative}}=q_i^\mu,\\
    &(J_S^{\mu})_{\mathrm{dissipative}}=\frac{q^{\mu}-\mu_iq_i^\mu}{\mathcal{T}},
\end{aligned}
\end{equation}
where $\zeta$ is the bulk viscosity and $\eta$ is the shear viscosity.
These first-order terms all contribute to the dissipative effects.
Also, we defined expansion $\vartheta$, shear tensor $\sigma^{\mu\nu}$, heat flux $q^\mu$ and diffusion currents $q^\mu_i$ as follows.
\begin{equation}
\begin{aligned}\label{eq: dissipation variables}
    \vartheta&=\nabla_{\mu}u^{\mu},\\
    \sigma^{\mu\nu}&=\frac{1}{2}\left(P^{\mu\lambda}\nabla_{\lambda}u^{\nu}+P^{\nu\lambda}\nabla_{\lambda}u^{\mu}\right)-\frac{1}{d-1}\vartheta P^{\mu\nu}, \\
    q^{\mu}&=-\kappa P^{\mu\nu}(\partial_{\nu}\mathcal{T}+\mathcal{T}u^{\rho}\nabla_{\rho}u_{\nu}),\\
    q_i^\mu&=-D_{ij}P^{\mu\nu}\partial_{\nu}\left(\frac{\mu_j}{\mathcal{T}}\right).
\end{aligned}
\end{equation}
where $\kappa$ is the thermal conductivity, and $D_{ij}$ are diffusion coefficients.
Note that these tensor and vectors are all orthogonal to $u^\mu$, hence all spatial.
$\sigma^{\mu\nu}, q^\mu, q^\mu_i$ being spatial is actually important, because from (see e.g. \cite{Andersson:2006nr})
\begin{align}
    \mathcal{T}\nabla_\mu J^\mu_S=
    \frac{q^\mu q_\mu}{\kappa \mathcal{T}}
    +\mathcal{T}(D^{-1})^{ij}q^\mu_i q_{j\mu}
    +\zeta \vartheta^2
    +2\eta \sigma^{\mu\nu}\sigma_{\mu\nu}\;,
\end{align}
we see that the entropy production is always non-negative due to the first-order effects, as we assume that the parameters $\kappa, D, \zeta, \eta$ are non-negative.

Once we have constitutive relations, equations of motion of fluid mechanics are given by
\begin{equation}
\begin{aligned}
    \nabla_\mu T^{\mu\nu}&=\partial_\mu T^{\mu\nu}
    +\Gamma^\mu_{\mu\lambda}T^{\lambda\nu}
    +\Gamma^\nu_{\mu\lambda}T^{\mu\lambda}=0\;, \\
    \nabla_\mu J^\mu_i&=\partial_\mu J^\mu_i
    +\Gamma^\mu_{\mu\lambda}J^\lambda_i=0\;.
\end{aligned}
\end{equation}

So far, we haven't imposed conformal invariance.
It turns out that it suffices to require the stress tensor to be traceless.\footnote{
In even dimensions, due to the Weyl anomaly, a trace of the stress tensor is proportional to linear combinations of appropriate powers of curvatures. But since the Riemann tensor has two derivatives in the background metric, it should contribute to the second order or higher in the derivative expansion. This holds true when the fluid solution's typical length scale is the same as the curvature radius.
}
The conditions we get are,
\begin{equation}
\begin{aligned}\label{eq: traceless condition}
    \mathcal{P}&=\frac{\epsilon}{d-1}\;,\\
    \zeta&=0\;.
\end{aligned}
\end{equation}

In fact, there is an ambiguity of field redefinition of $\mathcal{T},\mu_i,u^\mu$, which is absent at the zeroth order.
One way of fixing this ambiguity is to require
\begin{equation}
    \begin{aligned}
        u_{\mu}T^{\mu\nu}_{\mathrm{dissipative}}=0\;,
    \end{aligned}
\end{equation}
hence
\begin{equation}\label{eq: Landau frame}
    \begin{aligned}
        q^\mu=0\;.
    \end{aligned}
\end{equation}
This convention is called `Landau frame' and we will work on this convention from now on.

\subsubsection{Validity of fluid description}
Unless there exist some other protected operators that live as long as conserved currents (which will be the case for superfluid), the validity of fluid description translates to the validity of derivative expansion.

By looking at the equations \eqref{eq: perfect fluid constituitive}, \eqref{eq: first-order constitutive}, \eqref{eq: traceless condition}, and \eqref{eq: Landau frame}, we see that the derivative expansion at the first-order makes sense when
\begin{equation}
\begin{aligned}\label{eq: validity1}
    \eta  &\ll \epsilon L_{\mathrm{fluid}} \;,
\end{aligned}
\end{equation}
where $L_{\mathrm{fluid}}$ is the typical length scale of the fluid.
The quantity $\eta/\epsilon$ appears as the mean free path in the kinetic theory. Therefore, the condition \eqref{eq: validity1} is saying that the mean free path $l_{\mathrm{mfp}}$ should be small.

For holographic theories, it is known that \(\eta \approx \frac{s}{4\pi}\) \cite{Kovtun:2004de}. Therefore, the condition of $\eta\ll \epsilon L_{\mathrm{fluid}}$ is equivalent to $s\ll \epsilon L_{\mathrm{fluid}}$. 
For generic theories without known dual AdS, it was conjectured by \cite{Kovtun:2004de} that \(\eta \geq \frac{s}{4\pi}\). 
Therefore, $s\ll \epsilon L_{\mathrm{fluid}}$ is not a sufficient condition for the derivative expansion to be valid. However, for many known \emph{interacting} theories, \(\frac{\eta}{s}\) is not parametrically bigger than $\frac{1}{4\pi}$.\footnote{For the specific values of these ratios for various fluids, see \cite{Kovtun:2004de} for example.}

In weakly coupled theories, this ratio can be parametrically greater than the lower bound. For example in the $\lambda\phi^4$ theory in 4 dimensions, $\eta\sim \frac{s}{\lambda^2}$ \cite{Son:2007vk},
and the condition \eqref{eq: validity1} becomes $\frac{s}{\lambda^2}\ll \epsilon L_{\mathrm{fluid}}$.
We will see later that $\frac{\epsilon L_{\mathrm{fluid}}}{s}$ is proportional to the chemical potential for the solution we consider, and we take the chemical potential to be parametrically large. Thus the condition \eqref{eq: validity1} is satisfied when the chemical potential is much larger than $\frac{1}{\lambda^2}$. 

Therefore, we assume that the criterion $s\ll \epsilon L_{\mathrm{fluid}}$ can be applied generically—even for weakly coupled theories as long as we take the chemical potential sufficiently large—for the purpose of verifying the validity of derivative expansion.

\subsubsection{Extensivity}
On the other hand, conformal invariance constrains the grand canonical partition function of the static conformal fluid as follows:\footnote{For free theories, however, this does not hold, reflecting the fact that they are not well described by a fluid. Additionally, if a BPS moduli space exists in a theory, we encounter a similar problem. In this section, we only consider thermodynamic states and address the exceptional cases later.}
\begin{align}\label{eq: conformal fluid assumption}
    \frac{1}{V}\log Z_{gc}=\mathcal{T}^{d-1}h(\mu/\mathcal{T})=\mathcal{T}^{d-1}h(\nu),
\end{align}
where $V$ is volume of the space and $\nu$ is defined as the ratio between temperature and chemical potential $\nu\equiv \frac{\mu}{\mathcal{T}}.$
The thermodynamic potential of the static fluid is given as
$\Phi=\mathcal{E}-\mathcal{T}\mathcal{S}-\mu \mathcal{R}=-V\mathcal{T}^dh(\nu)$. Therefore, $h(\nu)$ is a function that determines the thermodynamics of the static fluid in AdS.
The first law of thermodynamics is written as
\begin{align}
    d\Phi=-\mathcal{S}d\mathcal{T}-\mathcal{P}d\mathcal{V}-\mathcal{R} d\mu.
\end{align}
Based on this, we obtain thermodynamic quantities as follows
\begin{align}\label{eq: charge relation}
    \rho &= (d-1)\mathcal{P}=(d-1)\mathcal{T}^{d}h(\nu),\nonumber\\
    J_i^t &=\mathcal{T}^{d-1}h_i(\nu),\nonumber\\
    s &=\mathcal{T}^{d-1}(dh-\nu_i h_i),
\end{align}
where $h_i(\nu)\equiv \frac{h(\nu)}{\partial \nu_i}$.

\subsection{Stationary solution on $\mathbb{R}\times S^{d-1}$}\label{subsection: stationary}

In this subsection, we extend the static solution \eqref{eq: charge relation} to find the solution for a stationary fluid, following the results of \cite{Bhattacharyya:2007vs}.
In this section, we usually set the radius of the sphere $R$ to be unity unless explicitly stated otherwise. Since $R$ was consistently traced in the previous section, it should be straightforward to recover $R$ by the dimensional analysis.

Our primary focus is on the system in equilibrium, where \( T_{\text{dissipative}} \) vanishes at order $\sim O\left(\frac{s}{\rho}\right)$. We do not address terms of order \( O\left(\frac{s^2}{\rho^2}\right) \).
For a stationary fluid on a sphere, the condition \( T^{\mu\nu}_{dissipative}=0 \) is equivalent to \(\sigma^{\mu\nu}=0\), \(q=0\), and \(q^i=0\).
First, \(\sigma^{\mu\nu} = 0\) requires the fluid motion to be a rigid rotation with angular velocities \(\Omega_a\), where \(a = 1, \dots, [d/2]\). Additionally, \(q = 0\) and \(q^i = 0\) require that \(\frac{\mathcal{T}}{\gamma} = T\) and \(\frac{\mu_i}{\mathcal{T}} = \nu_i\) are constants on the sphere.
The grand canonical partition function for the rotating fluid can be expressed by modifying the non-rotating fluid's partition function with a universal factor that depends on the angular velocities \cite{Bhattacharyya:2007vs}:
\begin{align}\label{partition_fluid 2}
\log Z_{rot} = \log \left[\text{Tr} \exp \left(\frac{1}{T}(-H + \Omega_a L_a + \zeta_i R_i)\right)\right] = \frac{\log Z_{non-rot}}{\prod_{a=1}^{[d/2]}(1 - \Omega_a^2)},
\end{align}
where \(L_a\) is the angular momentum, \(R_i\) is a global charge, \(\zeta_i\) is the chemical potential associated with \(R_i\), and \(\Omega_a\) is the chemical potential associated with \(L_a\). It is important to note that \(T\) and \(\mathcal{T}\) are distinct, as are \(\mu_i\) and \(\zeta_i\).

For concreteness, let $d=3$. We can choose the plane of rotation using an \(SO(3)\) transformation, so that \(u^{\mu}=\gamma(1,0,\Omega)\) and \(\gamma=(1-\Omega^2\sin^2\theta)^{-1/2}\).
Stress-energy tensor is given by \eqref{eq: perfect fluid constituitive} and non-vanishing components are explicitly written as
\begin{equation}
\begin{aligned}\label{eq: fluid stress tensor}
    T^{tt}&=T^3h(\nu)(3\gamma^{5}-\gamma^3), \\
    T^{t\phi}&=T^3h(\nu) 3\gamma^{5}\Omega,\\
    T^{\phi\phi}&=T^3h(\nu)(3\gamma^{5}\Omega^2+\gamma^3\csc^2\theta),\\
    T^{\theta\theta}&=T^3h(\nu)\gamma^3, 
\end{aligned}
\end{equation}
Note that $T^{t}_{t}+T^{\theta}_{\theta}+T^{\phi}_{\phi}=0,$ $\nabla_{\mu}T^{\mu\nu}=0$. Also, charge and entropy density are written as
\begin{align}\label{eq: fluid charge density}
    J_i^t=T^2\gamma^3 h_i(\nu),\qquad 
    J_S^t=T^2\gamma^3 (3h(\nu)-\nu_i h_i(\nu)).
\end{align}

The charges for the fluid are given by integrating charge density functions over the sphere, and written as follows.
\begin{equation}
\begin{aligned}\label{eq: charge relation_2}
    E&=\frac{2V_2 T^3 h(\nu)}{(1-\Omega^2)^2},\\
    S&=\frac{V_2 T^{2} [3h(\nu)-\nu_i h_i(\nu)]}{(1-\Omega^2)},\\
    J&=\frac{2V_2 T^3 h(\nu)\Omega}{(1-\Omega^2)^2}=E\Omega,\\
    R_i&=\frac{V_2 T^{2} h_i(\nu)}{(1-\Omega^2)}.
\end{aligned}
\end{equation}
where $V_2=Vol(S^2)=4\pi$ and $h_i=\frac{\partial h(\nu)}{\partial \nu_i}$.

\subsubsection{Zero temperature limit}
Suppose a zero temperature limit exists, as described by fluid dynamics. For the thermodynamic quantities in equation \eqref{eq: charge relation_2} to be well-defined in the \(T \rightarrow 0\) limit, the function \(h(\nu)\) should be expressed as follows:
\begin{align}
    h(\nu) = \alpha_1\nu^3 + \alpha_2\nu^2 + \cdots = \alpha_1\left(\frac{\zeta}{T}\right)^3 + \alpha_2\left(\frac{\zeta}{T}\right)^2 + \cdots.
\end{align}
where we assumed that there is only one internal charge $R_1\equiv Q$, and one corresponding chemical potential, $\zeta_1\equiv \zeta$.
Therefore, at \(T = 0\), the stress-energy tensor, charge density, and entropy density can be computed from equations \eqref{eq: fluid stress tensor} and \eqref{eq: fluid charge density}. Consequently, the charges and entropy can be written in terms of \(\zeta\) and \(\Omega\) as follows:
\footnote{
We could also consider the high-temperature limit, where $T\gg \zeta \gg 1$.
\begin{align}
    \Delta^2=J^2+\frac{1}{8\pi h(\nu)}(S+\nu Q)^3\approx J^2+\frac{1}{8\pi h(0)}S^3 
\end{align}
This relation is well known \cite{Bhattacharyya:2007vs,Shaghoulian:2015lcn,Benjamin:2023qsc}}
\begin{equation}
\begin{aligned}\label{eq: charge relation 2}
    E &= \frac{8\pi \alpha_1 \zeta^3}{(1-\Omega^2)^2},\\
    S &= \frac{8\pi \alpha_2 \zeta^2}{(1-\Omega^2)},\\
    J &= \frac{8\pi \alpha_1 \zeta^3 \Omega}{(1-\Omega^2)^2} = E\Omega,\\
    Q &= \frac{12\pi \alpha_1 \zeta^2}{(1-\Omega^2)}.
\end{aligned}
\end{equation}
The charge relation between energy and other conserved charges is expressed as:
\begin{align}\label{eq: sphere charge relation}
    \Delta^2 = J^2 + c_1^2 Q^{3} + \text{subleading terms},
\end{align}
where \( c_1 = \frac{1}{3\sqrt{3\pi \alpha_1}} \).
Given that the ratio between the leading and subleading terms is of the order \( l_{mfp}^2 \sim \left(\frac{S}{E}\right)^2 |_{\Omega=0} \sim \frac{1}{\zeta^2} \ll 1 \), the subleading term in \eqref{eq: sphere charge relation} is of the order \( O(\zeta^4) \). Therefore, for the relation \eqref{eq: sphere charge relation} to accurately capture the leading-order physics, we require \( J^2 \gg \zeta^4 \sim Q^2 \), meaning \( J \gg Q \).

In situations where the angular momentum is of order $O(1)$, a static fluid features a phonon mode.
The transition from phonon modes to collective modes occurs when the wavelength of the phonon becomes comparable to the mean free path of particles in the fluid.
In our system, the mean free path is of order $O(1/\zeta)$ and the wavelength of the excitation $\lambda \sim 1/J.$ Therefore, we observe phonon modes for $J\ll Q^{1/2}$.
Therefore we expect that the energy of the zero-temperature fluid given as
\begin{align}
     \Delta=c_1 Q^{3/2}+c_2 Q^{1/2} +\frac{1}{\sqrt{2}} \sqrt{J(J+1)} +\cdots .
\end{align}
$c_2$ is a constant determined by the large charge expansion of the static fluid.
For \(Q^{1/2} \ll J \ll Q\), subleading terms in \eqref{eq: sphere charge relation} dominate over \(J^2\). Therefore, the leading-order description is inaccurate.
The energy of the zero-temperature fluid can be expressed as:
\begin{align}
    \Delta = c_1 Q^{3/2} + f(Q,J) Q^{1/2} + \cdots,
\end{align}
where $f(Q,J)$ is a coefficient that depends on both \(Q\) and \(J\). Determining the exact value of $f(Q,J)$ necessitates a careful analysis of subleading effects, which we leave for future investigation.\footnote{We observe that this expansion is consistent with the single vortex–antivortex pair configuration that appears in a superfluid, as described \eqref{eq: vav energy}.}\\
For $Q\ll J\ll Q^{3/2}$ ($\frac{1}{\zeta}\ll\Omega\ll 1$), the equation \eqref{eq: sphere charge relation} can be rewritten as
\begin{align}\label{eq: fluid rigid rot 2}
    \Delta=c_1 Q^{3/2}+\frac{J^2}{2c_1 Q^{3/2}}+\cdots.
\end{align}
For $Q^{3/2}\ll J\ll Q^{2}$ ($\Omega\lessapprox 1$), the charge relation is written as 
\begin{align}\label{eq: fast rigid rot}
    \Delta=J+\frac{c_1^2}{2}\frac{Q^3}{J}+\cdots.
\end{align}

\subsection{Holographic fluids and AdS/EFT}\label{sec: AdS EFT}
For holographic theories, large AdS$_{d+1}$ black holes (with event horizon radius \( r_+ \) much larger than \( l_{\text{AdS}} \)) can be effectively described as fluids on \( S^{d-1} \times \mathbb{R} \) \cite{Bhattacharyya:2007vs}. The condition for a large event horizon corresponds to the fluid’s mean free path being small. This relationship arises because the mass of the black hole scales as \( r_+^d \), while the entropy scales as \( r_+^{d-1} \), leading to the mean free path of the fluid being \( l_{\text{mfp}} \sim \frac{S}{E} \sim \frac{1}{r_+} \ll 1 \).
In this subsection, we compute the stress-energy tensor and charge density of large-charge extremal black holes and compare these results with the fluid dynamics calculations presented in \S \ref{subsection: stationary}, establishing their agreement. We then discuss the similarities and differences between the zero-temperature fluid and the superfluid ground state identified in \S \ref{sec: superfluid}, particularly from the perspective of holography. 

\subsubsection{AdS$_4$ Kerr-Newman Black hole}
As a concrete example, we consider AdS$_4$ Kerr-Newman black hole with an electric charge $Q$ and angular momentum $J$, which is given as a solution of the Einstein-Maxwell equation \cite{Caldarelli:1999xj}
\footnote{While we focus on this particular case, the identical analysis can be applied to general non-supersymmetric black holes.}.
The metric of the AdS$_4$ Kerr-Newman black hole is given as
\begin{align}
    ds^2&=-\frac{(r^2+a^2)(1+r^2)-2mr+q^2}{r^2+a^2\cos^2\theta}\left[dt-\frac{a\sin^2\theta}{1-a^2}d\phi\right]^2+\frac{r^2+a^2\cos^2\theta}{(r^2+a^2)(1+r^2)-2mr+q^2}dr^2\nonumber\\
    &+\frac{r^2+a^2\cos^2\theta}{1-a^2\cos^2\theta}d\theta^2+\frac{(1-a^2\cos^2\theta)\sin^2\theta}{r^2+a^2\cos^2\theta}\left[adt-\frac{r^2+a^2}{1-a^2}d\phi\right]^2,
\end{align}
and the vector field is given as
\begin{align}
    A=-\frac{qr}{(r^2+a^2\cos^2\theta)^2}(dt-\frac{a\sin^2\theta}{1-a^2}d\phi).
\end{align}
The solution is determined by three parameters $m,q,$ and $-1<a<1$.

In order to obtain stress-energy tensor and charge density of the boundary theory, we transform the metric to make it manifestly asymptotically AdS$_4$ by the following transformation \cite{Henneaux:1985tv}
\begin{equation}
\begin{aligned}
    \phi'=(1-a^2)\phi+at,\quad t'=t\\
    r'\cos\theta'=r\cos\theta\\
    (1-a^2)r'^{2}=r^2+a^2\sin^2\theta-a^2r^2\cos^2\theta.
\end{aligned}
\end{equation}
The transformed metric is written as 
\begin{align}
    ds^2=ds_0^2+h_{\mu\nu}dx^{\mu}dx^{\nu},
\end{align}
where $ds_0^2$ is the metric of the global AdS$_4$ and $h_{\mu\nu}$ is given by
\begin{equation}\begin{split}
    h_{tt}&=\frac{2m}{r}(1-a^2\sin^2\theta)^{-5/2}+O(r^{-3}),\\
    h_{t\phi}&=-\frac{2ma\sin^2\theta}{r}(1-a^2\sin^2\theta)^{-5/2}+O(r^{-3})\\
    h_{\phi\phi}&=\frac{2ma^2\sin^4\theta}{r}(1-a^2\sin^2\theta)^{-5/2}+O(r^{-3})\\
    h_{rr}&=\frac{2m}{r^5}(1-a^2\sin^2\theta)^{-3/2}+O(r^{-7})\\
    h_{\theta r}&=-\frac{2ma^2}{r^4}(1-a^2\sin^2\theta)^{-5/2}\sin\theta\cos\theta+O(r^{-6})\\
    h_{\theta\theta}&=\frac{2ma^4}{r^3}(1-a^2\sin^2\theta)^{-7/2}\sin^2\theta\cos^2\theta+O(r^{-5})
    \end{split}
\end{equation}
From this, one can obtain the stress-energy tensor of the boundary state: 
\begin{equation}\label{eq: BH stress}\begin{split}
		&8\pi G T^{t}_{t}=-\frac{m \left( a^2 \sin ^2{\theta }+2\right)}{\left(1-a^2 \sin^2 \theta \right)^{5/2}}\\
		&8\pi G T^{\phi}_{\phi}=\frac{ m \left(1+2 a^2\sin^2\theta\right)}{\left( 1-a^2 \sin^2 \theta \right)^{5/2}}\\
		&8\pi G T^{t}_{\phi}=\frac{3 m a\sin^2\theta }{\left(1-a^2 \sin ^2{\theta}\right)^{5/2}}\\
		&8\pi G T^{\theta}_{\theta}=\frac{ m}{\left(1-a^2 \sin ^2{\theta}\right)^{3/2}},
	\end{split}	
\end{equation}
and similarly the charge density:
\begin{align}\label{eq: BH charge density}
    \rho=\frac{1}{4\pi}\frac{q}{(1-a^2\sin^2\theta)^{3/2}}.
\end{align}

Thermodynamic quantities of the black hole are given as
\begin{equation}
\begin{aligned}
    E=\frac{m}{(1-a^2)^2}, \qquad J=\frac{ma}{(1-a^2)^2}, \qquad Q=\frac{q}{1-a^2},\\
    \beta=\frac{4\pi(r_+^2+a^2)}{r_+(1+a^2+3r_+^2-(a^2+q^2)/r_+^2)},\qquad S=4\pi \frac{r_+^2+a^2}{1-a^2}.
\end{aligned}
\end{equation}
where the radius of event horizon $r_+$ is given as the largest root of the equation $\Delta_{r}=(r^2+a^2)(1+r^2)-2mr+q^2=0.$
Also, angular velocity $\Omega$ and electric potential $\zeta$ are given as
\begin{align}
    \Omega=\frac{a(1+r_+^2)}{r_+^2+a^2}, \qquad     \zeta = \frac{q r_+}{r_+^2+a^2}.
\end{align}

Now, we determine the regime of charges for black hole states that can be described by fluid dynamics. The mean free path \( l_{\text{mfp}} \) is given by
\begin{align}
    l_{\text{mfp}} \sim \frac{S}{4\pi E} \bigg|_{a=0} = \frac{r_+^2}{2r_+^3 + r_+}.
\end{align}
Thus, taking the fluid dynamics limit corresponds to taking the horizon radius much larger than the radius of AdS$_4$, i.e., \( r_+ \rightarrow \infty \). In this limit, $\Omega \approx a$ and $\zeta \approx \frac{q}{r_+}.$
We are particularly interested in the extremal limit \( T \rightarrow 0 \) because we want to find the charge relation of the lowest-energy state. The extremality condition is given by
\begin{align}
    q^2 = 3r_+^4 + (1 + a^2)r_+^2 + a^2.
\end{align}

The thermodynamic potential at zero temperature, \(\Phi = E - Q\zeta - \Omega J = -\frac{1}{\beta}\log Z\), can be expressed in terms of \((r_+, a)\) or \((\Omega, \zeta)\):
\begin{align}\label{eq: thpot ext BH}
    \Phi = -\frac{r_+^5}{(r_+^2 + a^2)(1 - a^2)} = \frac{1}{3\sqrt{3}}\left(-\frac{\zeta^3}{1 - \Omega^2} + \frac{3\zeta(1 + \Omega^2)}{2(1 - \Omega^2)} + \cdots\right).
\end{align}
In the fluid dynamics limit, which corresponds to large \(r_+\) (or large \(\zeta\)) limit, the leading term is given by \(\Phi = -\frac{1}{3\sqrt{3}}\frac{\zeta^3}{1 - \Omega^2}\). This result is consistent with \eqref{eq: conformal fluid assumption} and \eqref{partition_fluid 2}.

For zero angular momentum, $a=\Omega=0$, and a black hole is described by a static fluid
\begin{equation}
\begin{aligned}
    E=2r_+^3+r_+,\\
    Q=\sqrt{3r_+^4+r_+^2},
\end{aligned}
\end{equation}
and the charge relation is written as
\begin{equation}
\begin{aligned}\label{eq: static}
    E=\frac{\sqrt{\sqrt{12 Q^2+1}-1} \left(\sqrt{12 Q^2+1}+2\right)}{3 \sqrt{6}}=
    c_1 Q^{3/2}+c_2 Q^{1/2}+ \cdots,
\end{aligned}
\end{equation}
where $c_1=\frac{2}{3^{3/4}}, c_2=\frac{1}{2\cdot 3^{1/4}}$.

For $Q^{1}\ll J\ll Q^{2},$ energy and charge of the black hole are consistent with that of the stationary rotating fluids given in \eqref{eq: charge relation}.
\begin{align}
    E=\frac{2r_+^3}{(1-\Omega^2)^2}+\cdots,\qquad 
    J=\frac{2r_+^3\Omega}{(1-\Omega^2)^2}+\cdots,\qquad
    Q=\frac{\sqrt{3}r_+^2}{1-\Omega^2}+\cdots.
\end{align}
The charge relation is given by
\begin{align}
    E^2=J^2+ c_1^2 Q^{3}+\text{subleading}.
\end{align}

\subsubsection{AdS/EFT}\label{adseft}
The authors of \cite{Loukas:2018zjh} made the interesting observation that the energy-charge relationship of the extremal \( \text{AdS}_4 \)-RN black hole, as given in equation \eqref{eq: static}, precisely matches that of the classical ground state in the \( O(2) \) linear sigma model on \( S^2 \times \mathbb{R} \) with fixed charge $Q$, whose Lagrangian is expressed as
\begin{align}
    L=-\partial_{\mu}\varphi \partial^{\mu}\varphi^*-\frac{\mathfrak{R}}{8}\varphi\varphi^*-\frac{\lambda}{3}(\varphi\varphi^*)^3
\end{align}
where $\mathfrak{R}=\frac{2}{R^2}$ is the scalar curvature of the sphere with radius $R$. This model is classically conformal invariant.
The ground state energy at charge $Q$ is given by\footnote{The details of the derivation is in Appendix \ref{appendix B}.}
\begin{align}
    E=\frac{\pi}{3\sqrt{2}R}\frac{1}{\sqrt{\lambda}}
    \left(2+\sqrt{1+\frac{4\lambda}{\pi^2}Q^2}\right)
    \sqrt{\sqrt{1+\frac{4\lambda}{\pi^2}Q^2}-1}. \label{gsenergy}
\end{align}
The functional form of equation \eqref{eq: static} and \eqref{gsenergy} are exactly the same.

In fact, this correspondence can be extended to arbitrary dimensions, establishing a connection between extremal \( \text{AdS}_{d+1} \)-RN black holes and ground states of the linear sigma model on \( S^{d-1} \times \mathbb{R} \). Further details on the derivation of the correspondence are provided in Appendix \ref{appendix C}. 
The agreement between the energy-charge relations in extremal AdS black holes and the superfluid ground state is referred to as the AdS/EFT correspondence \cite{Loukas:2018zjh, Liu:2020uaz, delaFuente:2020yua}. At large charge, this correspondence is intuitive at leading order: large AdS black holes can be described by fluid dynamics, while the energy of the superfluid ground state is effectively described by a zero-temperature fluid.

Even in scenarios involving large angular momentum  $J$, the presence of fluid-like descriptions in both theories does not invalidate the AdS/EFT correspondence.
Further evidence supporting AdS/EFT at large spin comes from the matching stress-energy tensor and charge density of the AdS Kerr-Newman black hole \eqref{eq: BH stress}, \eqref{eq: BH charge density} with those of the ground state of the $O(2)$ linear sigma model at leading order \eqref{eq: O2 stress charge}.

Although the AdS/EFT correspondence appears appealing, it is important to recognize fundamental differences between the superfluid (the ground state of the EFT) and the zero-temperature normal fluid (the AdS extremal black hole). The former possesses a nearly unique ground state, whereas the latter exhibits a macroscopic number of states. Consequently, there has been speculation that the superfluid phase might correspond to a geometry without a horizon in holographic theories. For example, as suggested in \cite{Liu:2020uaz}, boson stars—which reduce to holographic superconductors in the thermodynamic limit—emerge as potential candidates for the dual of the superfluid. Furthermore, whether this correspondence holds at all orders in the large charge and large angular momentum expansions beyond the leading approximation, particularly in the presence of quantum effects, remains an open question for future investigation.

Nonetheless, it is also known that in the extreme zero-temperature limit of an extremal black hole, the density of states vanishes due to the so-called Schwarzian mode living at the boundary of the AdS$_2$ throat \cite{Iliesiu:2020qvm}. Specifically, when quantum effects are accounted for, the density of states should vanish even for black hole states (and their corresponding fluid states). This observation raises the possibility that the EFT might, albeit indirectly, capture certain aspects of extremal black hole physics.  

That said, the near-ground state spectrum of the black hole remains dense, with an average energy gap of order $\sim e^{-S_0}$, where $S_0$ is the classical extremal entropy—giving the appearance of a continuous spectrum. In contrast, the spectrum in the EFT is sparse. Thus, while some aspects of the correspondence might hold, the connection remains poorly understood, albeit for somewhat different reasons.


\section{Conclusions and Future directions}\label{sec: conc}
In this paper, we investigate the ground-state energy spectrum of CFTs on $S^2$ with large charge and angular momentum. The system is governed by an EFT formulated in terms of a scalar field $\chi$.

In the large angular momentum regime $Q \ll J\ll Q^2$, the system generates densely populated vortices, introducing vortex currents as emergent degrees of freedom in addition to $\chi$. In \S \ref{sec: superfluid}, we solve the EFT using these three fundamental degrees of freedom ($\chi+$ two components of the vortex current) and establish the universal behavior of the energy $E$ across various parametric regimes of charge $Q$ and angular momentum $J$. Specifically, we show that the large-charge expansion of the energy has the same functional form of the energy-charge relationship of a zero-temperature fluid state, as derived in \eqref{eq: superfluid charges} and \eqref{eq: charge relation 2}.

It is crucial to emphasize again that the similarity in the functional forms of the energy and charge density between the stationary superfluid and the zero-temperature fluid does not imply identical microscopic dynamics. A simple example is the $J = 0$ case, where the superfluid exhibits a unique ground state. Conversely, the zero-temperature fluid in a holographic theory corresponds to an extremal black hole, which supports a macroscopic number of states \footnote{Note that when the temperature is of order \(O\left(\frac{1}{Q^{1/2}}\right) \sim O\left(\frac{1}{r_+}\right)\), the quantum effects (Schwarzian) become significant, causing the density of states to vanish \cite{Iliesiu:2020qvm}. However, in the leading-order fluid dynamics description, subleading terms of order \(\delta E \sim O(r_+)\) are neglected. This implies that only temperatures of order \(T \sim O(1)\), much larger than the temperature scale where quantum effects become significant (\(T \sim O\left(\frac{1}{r_+}\right)\)), are resolved. As a result, the fluid dynamics description at the leading order does not capture quantum effects.}
Furthermore, the quasi-normal modes in the black hole background are dissipative, while the phonon modes in the superfluid background are non-dissipative. 
It is therefore interesting to study the subleading order behavior, including quantum effects, to see whether there is a characteristic that distinguishes zero-temperature fluid from the superfluid phase.

Another interesting question is how black hole states, which exhibit extensive entropy, can be described from the perspective of EFT. One possibility is the existence of metastable macroscopic states above the ground state, which could provide a description of a fluid-like phase. To explore this possibility, it might be necessary to incorporate additional terms—such as matter contributions—into the effective action proposed in \eqref{eq: effective action}. However, we do not have any concrete proposals at this stage.

Furthermore, we have not conducted an exhaustive analysis of the fluctuations and full dynamics of the fast rigidly rotating phase. In particular, we have not addressed the vortex lattice fluctuation modes, which arise due to vortex-vortex interactions and self-energy contributions \cite{Moroz:2018noc,Cuomo:2021qws}. We believe that these fluctuation modes could emerge by including one-derivative terms in the effective action for $\xi$. Notably, the subleading corrections to the chiral mode might be influenced by vortex lattice fluctuations, potentially leading to mixing effects. A detailed investigation into these emergent dynamics, particularly the possible connection to fluid dynamics arising from vortex density fluctuations would be an intriguing direction for future study.

In the paper, we exclusively studied $d=3$ with $U(1)$ global symmetry. Here we briefly comment on the ground state of the higher-dimensional CFTs with non-Abelian global symmetry.
We expect that the ground state should be effectively described by a zero-temperature fluid state in general dimensions, where the energy and angular momenta are given by (See, e.g. \cite{Bhattacharyya:2007vs} for the expression at the finite temperature):  
\begin{equation}
\begin{split}
    E&=\frac{V_{d-1} h(\mu)}{\prod_{b}(1-\Omega_b^2)}\left[2\sum_a\frac{\Omega_a^2}{1-\Omega_a^2}+d-1\right],\\
    J_a&=\frac{V_{d-1} h(\mu)}{\prod_{b}(1-\Omega_b^2)}\left[\frac{2\Omega_a}{1-\Omega_a^2}\right],~~~a=\left(1,\cdots,\left[\frac{d}{2}\right]\right),\\
    R_i&=\frac{V_{d-1} h_i(\mu)}{\prod_{b}(1-\Omega_b^2)}
    \end{split}
\end{equation}
where $V_{d-1}=\text{Vol}(S^{d-1})=\frac{2\pi^{d/2}}{\Gamma(d/2)}$.
$h(\mu)$ is a function of $\mu_i$s, homogeneous polynomial of degree $d$.
$h(\mu_i)=\frac{\partial h(\mu)}{\partial \mu_i}$.
As an illustration, let us consider the case of $d=4$ with a $U(1)$ charge $Q$ and two angular momenta $J_1$ and $J_2$.
\begin{align}
\begin{split}
    E&=\frac{V_3 \alpha \mu^4}{(1-\Omega_1^2)(1-\Omega_2^2)}\left[2\frac{\Omega_1^2}{1-\Omega_1^2}+2\frac{\Omega_2^2}{1-\Omega_2^2}+3\right],\\
    J_a&=\frac{V_3 \alpha \mu^4}{(1-\Omega_1^2)(1-\Omega_2^2)}\left[\frac{2\Omega_a}{1-\Omega_a^2}\right],~~~a=(1,2),\\
    Q&=\frac{4V_3 \alpha \mu^3}{(1-\Omega_1^2)(1-\Omega_2^2)},
    \end{split}
\end{align}
\footnote{ Here we see that $J_1J_2\ll Q^3$ as long as $\mu\gg 1$. Also, each angular momentum: $J_a\ll Q^2$.}
where $V_3=2\pi^2$.
For a zero angular momenta state, we obtain
\begin{align}
    E=c_1 Q^{4/3}+c_2 Q^{2/3}+\cdots
\end{align}
where $c_1=\frac{3}{(4V_3\alpha)^{1/3}}$.

Below, we identify key parametric regimes of \( J_1 \) and \( J_2 \) where the energy simplifies. The results depend on the angular velocities \(\Omega_1\) and \(\Omega_2\), and we provide the corresponding conditions under which the expansions are valid:

\begin{enumerate}
    \item For small angular velocities (\(\Omega_1, \Omega_2 \ll 1\)):  
    In this regime, the energy expansion is:  
    \begin{align}
        E \approx c_1 Q^{4/3} + \frac{3}{4c_1} \frac{J_1^2 + J_2^2}{Q^{4/3}}.
    \end{align}  
    Here, the leading term is proportional to \( Q^{4/3} \), and the subleading correction involves \( J_1^2 + J_2^2 \). The expansion is valid when the subleading term is larger than \( Q^{2/3} \), which requires \( J_a \gg Q \) for \( a = 1, 2 \).  

    \item For one angular velocity close to unity (\(\Omega_1 \sim 1, \Omega_2 \ll 1\)):  
    When \(\Omega_1\) approaches unity, the energy becomes:  
    \begin{align}
        E \approx J_1 + \left(\frac{2c_1}{3}\right)^{\frac{3}{2}} \frac{Q^2}{J_1^{1/2}}.
    \end{align}  
    In this case, the dominant contribution is linear in \( J_1 \), with a subleading correction proportional to \( Q^2 / J_1^{1/2} \). The validity of this expansion requires \( Q^{4/3} \ll J_1 \ll Q^2 \) and \( Q \ll J_2 \ll Q^{4/3} \).  

    \item For both angular velocities near unity (\(\Omega_1, \Omega_2 \sim 1\)):  
    When both angular velocities approach unity, the energy simplifies to:  
    \begin{align}
        E \approx J_1 + J_2 + \frac{c_1^3}{27} \frac{Q^4}{J_1 J_2}.
    \end{align}  
    The leading terms are linear in \( J_1 \) and \( J_2 \), with a subleading correction proportional to \( Q^4 / (J_1 J_2) \). This expansion is valid in the regime \( Q^{8/3} \ll J_1 J_2 \ll Q^3 \).  
\end{enumerate}  
In each case, the energy captures the interplay between the charge \( Q \) and the angular momenta \( J_1, J_2 \), with the parametric regimes ensuring the validity of the approximations.
For $d \geq 4$, the ground state is expected to correspond to a densely populated higher-dimensional generalization of vortices, such as vortex strings in $d = 4$ \cite{Horn:2015zna,Cuomo:2019ejv}. Exploring the effective field theory of higher-dimensional defects would be an interesting future direction. It is possible that such configurations exhibit intriguing topological structures that warrant further investigation.

Also, another promising avenue involves studying CFTs with moduli spaces \cite{Hellerman:2015nra,Hellerman:2017veg,Jafferis:2017zna,Cuomo:2024fuy}, such as supersymmetric theories, and examining how the addition of angular momentum impacts these setups. 
As explained briefly in the introduction, the ground state at $J=0$ does not follow the large charge expansion of the energy $\Delta \approx c_1 Q^{3/2},$ but $\Delta\approx Q$.
In the presence of angular momentum. It is presumably that the ground state energy is given by
\begin{align}
    \Delta=Q+J+\text{subleading}.
\end{align}
that almost saturates the unitarity bound.
It would be interesting to see how angular momenta influence the structure of the moduli space and its near-ground state spectrum and whether they give rise to novel phenomena.

Finally, we would like to mention that in \cite{Cuomo:2022kio}, the authors identified a giant vortex solution within the EFT, which appears to exist as a metastable phase. This finding suggests the presence of a rich landscape of metastable states, potentially sharing similar properties with the giant vortex and the rigid rotation phase. Investigating the stability criteria and possible transitions between these states could provide deeper insights into the structure of the EFT. For instance, it would be interesting to determine the precise conditions under which such metastable configurations decay and whether these processes can be characterized universally within the EFT framework.

\section*{Acknowledgments}
We would like to thank Seok Kim, Zohar Komargodski, Shiraz Minwalla, Masataka Watanabe, and especially Gabriel Cuomo for very useful comments and discussions. 
The work of J.C. was supported by the NRF grant  2021R1A2C2012350. The work of E.L. was supported by Basic Science Research Program through the National Research Foundation of Korea (NRF) funded by the Ministry of Education RS-2024-00405516 and the Infosys Endowment for the study of the Quantum Structure of Spacetime.
\appendix

\section{Linear sigma model}\label{appendix B}

In this Appendix, we investigate the large charge expansion of a CFT, whose ground state is a superfluid state.
As one of the simplest examples, let us consider a $O(2)$ linear sigma model, whose Lagrangian is written as
\begin{align}
    L=-\partial_{\mu}\varphi \partial^{\mu}\varphi^*-\frac{\mathfrak{R}}{8}\varphi\varphi^*-\frac{\lambda}{3}(\varphi\varphi^*)^3
\end{align}
where $\mathfrak{R}=\frac{2}{R^2}$ is the scalar curvature of the sphere with radius $R$. This model is classically conformal invariant.
The equation of motion is written as
\begin{align}
    \frac{1}{\sqrt{-g}}\partial_{\mu}\sqrt{-g}\partial^{\mu}\varphi-\frac{\mathfrak{R}}{8}\varphi-\lambda(\varphi\varphi^*)^2\varphi=0,
\end{align}
or equivalently,
\begin{align}
    \lambda(\varphi\varphi^*)^3=\varphi^*\nabla_{\mu}\nabla^{\mu}\varphi-\frac{\mathfrak{R}}{8}\varphi\varphi^*.
\end{align}

We now assume the ground state with charge $Q$ to be homogeneous and static.
It is convenient to rewrite the complex scalar as $\varphi\equiv\phi_1+i\phi_2\equiv ue^{i\alpha}$ where $u,\alpha$ are real. Then, $\frac{Q}{4\pi R^2}=\rho=\phi_1\dot{\phi_2}-\phi_2\dot{\phi_1}=\frac{1}{2i}(\varphi^*\dot{\varphi}-\dot{\varphi}^*\varphi)=u^2\dot{\alpha}$.
The equation of motion is written as
\begin{align}
    &u^2\dot{\alpha}=\rho,\\
    &\ddot{u}=u\dot{\alpha}^2-\frac{\mathfrak{R}}{8}u-\lambda u^5.
\end{align}
Plugging $\dot{\alpha}=\rho/u^2$ into the second equation,
\begin{align}
    \ddot{u}&=\frac{\rho^2}{u^3}-\frac{\mathfrak{R}}{8}u-\lambda u^5
    \equiv -\frac{dV_{\mathrm{eff}}}{du},\\
    V_{\mathrm{eff}}&=\frac{\rho^2}{2u^2}+\frac{\mathfrak{R}}{16}u^2+\frac{\lambda}{6}u^6+(\mathrm{const.}).
\end{align}
From this, we see that the classical ground state is given by
\begin{align}
    \alpha=\mu t+(\mathrm{const.}),\;\;
    u=u_0,
\end{align}
where $\mu,u_0$ are constants and determined by (in terms of $Q=4\pi R^2\rho$)
\begin{align}
    \mu=\rho/u_0^2,\;\;
    \lambda u_0^4=\mu^2-\frac{\mathfrak{R}}{8}.
\end{align}
from which we get
\begin{align}
    u_0^4=\frac{\mathfrak{R}}{16\lambda}\left(
    \sqrt{1+4\lambda \rho^2\left(\frac{8}{\mathfrak{R}}\right)^2}-1
    \right).
\end{align}
Then, the ground state energy is,
\begin{align}
    E&=4\pi R^2\left(\dot{u}^2+u^2\dot{\alpha}^2
    +\frac{\mathfrak{R}}{8}u^2+\frac{\lambda}{3}u^6\right)\bigg|_{u=u_0,\dot\alpha^2=\mu^2=\lambda u_0^4+\frac{\mathfrak{R}}{8}}
    \nonumber\\
    &=4\pi R^2\left(\frac{\mathfrak{R}}{4}u_0^2+\frac{4\lambda}{3}u_0^6\right)
    =4\pi R^2 \frac{4}{3}\left(\lambda u_0^4+\frac{3}{16}\mathfrak{R}\right)u_0^2
    \nonumber\\
    &=\frac{\pi}{3\sqrt{2}R}\frac{1}{\sqrt{\lambda}}
    \left(2+\sqrt{1+\frac{4\lambda}{\pi^2}Q^2}\right)
    \sqrt{\sqrt{1+\frac{4\lambda}{\pi^2}Q^2}-1} \nonumber\\
    &=c_1 Q^{3/2}+c_2 Q^{1/2}+\cdots 
\end{align}
where $c_1=\frac{2\lambda^{1/4}}{3\sqrt{\pi}}$.

\subsection{Rigid rotation phase}\label{linsigrig}
We consider the following ansatz for the rigid rotation phase
\begin{align}
    \varphi = u \text{Pexp}\left[\int if(\theta)(dt-\Omega \sin^2\theta d\phi)\right].
\end{align}
The equation of motion is written as
\begin{align}
    \lambda |u|^4=f^2(\theta)(1-\Omega^2\sin^2\theta),
\end{align}
because
\begin{align}
    \nabla_{\rho}\nabla^{\rho}\varphi&=\left[-\partial_t^2+\frac{1}{\sin^2\theta}\partial_{\phi}^2+\partial_{\theta}^2+\cot\theta\partial_{\theta}\right] \varphi\nonumber\\
    &\approx f^2(\theta)\left(1-\Omega^2\sin^2\theta\right)\varphi.
\end{align}
where we neglect the terms containing $\partial_{\theta}$ because $f(\theta)$ is a large parameter.
The following equations solve the EOM and minimize the total energy.
\begin{align}
    f=\frac{\mu}{1-\Omega^2\sin^2\theta} \quad \text{and} \quad \lambda|u|^4=\frac{\mu^2}{1-\Omega^2\sin^2\theta}.
\end{align}
The stress-energy tensor of the theory is written as
\begin{align}
    T_{\mu\nu}
    &=2\partial_{\{\mu}\varphi^*\partial_{\nu\}}\varphi
    -g_{\mu\nu}\bigg[
    \partial^{\mu}\varphi^*\partial_{\mu}\varphi
    +\frac{R}{8}\varphi^*\varphi
    +\frac{\lambda}{3}(\varphi^*\varphi)^3
    \bigg]
    \nonumber\\
    &\quad+\frac{1}{4}(R_{\mu\nu}
    +g_{\mu\nu}\nabla^2-\nabla_\mu\nabla_\nu)(\varphi^*\varphi)
\end{align}
where $R_{\mu\nu}=\mathrm{diag}(0,1,\sin^2\theta)$ is the Ricci tensor of the sphere.
The stress-energy tensor and charge density at the leading order are given as
\begin{equation}
\begin{aligned}\label{eq: O2 stress charge}
    T_{tt}=\frac{2}{\sqrt{\lambda}}\frac{\mu^3(2+\Omega^2\sin^2\theta)}{3(1-\Omega^2\sin^2\theta)^{5/2}}, \\
    T_{t\phi}=-\frac{2}{\sqrt{\lambda}}\frac{\mu^3\Omega\sin^2\theta}{(1-\Omega^2\sin^2\theta)^{5/2}},\\
    T_{\phi\phi}=\frac{2}{\sqrt{\lambda}}\frac{\mu^3(\sin^2\theta+2\Omega^2\sin^4\theta)}{3(1-\Omega^2\sin^2\theta)^{5/2}},\\
    T_{\theta\theta}=\frac{2}{\sqrt{\lambda}}\frac{\mu^3}{3(1-\Omega^2\sin^2\theta)^{3/2}},\\
    \rho=\frac{2}{\sqrt{\lambda}}\frac{\mu^2}{(1-\Omega^2\sin^2\theta)^{3/2}},
\end{aligned}
\end{equation}
thereby reproducing \eqref{eq: superfluid charges} and the large charge expansion of the energy $E^2-J^2=c_1^2Q^3$.


\section{AdS/EFT correspondence for general dimensions}\label{appendix C}

In this section, we demonstrate that the AdS/EFT correspondence holds not only for \(d=3\) but also for general dimensions with \(d > 2\). First, we identify the homogeneous ground state of the \(O(2)\) model on \(S^{d-1} \times \mathbb{R}\) for a fixed charge and compare its energy to that of the extremal AdS\(_{d+1}\)-RN black hole with the same charge.

The Lagrangian of an \(O(2)\) model on \(S^{d-1} \times \mathbb{R}\) is given by
\begin{align}
    \mathcal{L} = -\partial_{\mu}\varphi \partial^{\mu} \varphi^{*} - \frac{(d-2)R}{4(d-1)} \varphi \varphi^{*} - \frac{d-2}{d} \lambda (\varphi \varphi^{*})^{\frac{d}{d-2}},
\end{align}
where \(R\) is the curvature scalar of \(S^{d-1}\).
The equation of motion is
\begin{align}
    \lambda (\varphi \varphi^{*})^{\frac{d}{d-2}} = \varphi^{*} \nabla_{\mu} \nabla^{\mu} \varphi - \frac{(d-2)R}{4(d-1)} \varphi \varphi^{*}.
\end{align}
We employ the following homogeneous ansatz: \(\varphi = u e^{i \mu t}\). This reduces the equation of motion to
\begin{align}
    \lambda u^{\frac{2d}{d-2}} = u^2 \mu^2 - \frac{(d-2)R}{4(d-1)} u^2,
\end{align}
which provides the relationship between \(u\) and \(\mu\).
The charge density \(\rho\) and energy density \(\epsilon\) are then given by
\begin{align}\label{eq: CFTd lce}
    \rho &= \frac{1}{2i} (\varphi \dot{\varphi}^{*} - \varphi^{*} \dot{\varphi}) = u^2 \mu = u^2 \sqrt{\lambda u^{\frac{4}{d-2}} + \frac{(d-2)R}{4(d-1)}}, \nonumber\\
    \epsilon &= u^2 \mu^2 + \frac{(d-2)R}{4(d-1)} u^2 + \frac{d-2}{d} \lambda u^{\frac{2d}{d-2}} = \frac{(d-2)R}{2d(d-1)} u^2 + \frac{2d-2}{d} \lambda u^{\frac{2d}{d-2}}.
\end{align}

Let us consider the Reissner-Nordström-AdS\(_{d+1}\) (RN-AdS\(_{d+1}\)) black hole solution, (as discussed in \cite{Guo:2015swu}, for example), with the action given by
\begin{align}
    \frac{1}{16\pi G_{d+1}} \int d^{d+1}x \sqrt{-g} \left(R + \frac{d(d-1)}{L^2} - \frac{L^2}{g_s^2} F_{\mu\nu} F^{\mu\nu}\right).
\end{align}
The metric of the black hole is
\begin{align}
    ds^2 = -f(r) dt^2 + \frac{dr^2}{f(r)} + r^2 d\Omega_{d-1}^2,\nonumber\\
    A = a\left(1 - \frac{r_+^{d-2}}{r^{d-2}}\right) dt,
\end{align}
where
\begin{align}
    f(r) &= 1 - \frac{M}{r^{d-2}} + \frac{q^2}{r^{2d-4}} + \frac{r^2}{L^2}, \\
    a &= \sqrt{\frac{d-1}{2(d-2)}} \frac{g_s q}{L r_+^{d-2}}.
\end{align}
The mass \(E\) and charge \(Q\) are given by
\begin{align}
    E &= \frac{(d-1)M \Omega_{d-1}}{16\pi G_{d+1}}, \nonumber\\
    Q &= \frac{\sqrt{2(d-1)(d-2)} q \Omega_{d-1}}{8\pi G_{d+1} g_s}.
\end{align}
The radius of the event horizon, \(r_+\), is determined by the equation \(f(r_+) = 0\):
\begin{align}
    M = r_+^{d-2} + \frac{q^2}{r_+^{d-2}} + \frac{r_+^d}{L^2}.
\end{align}
The extremality condition is given by:
\begin{align}
    1 - \frac{d-2}{d} \frac{L^2 (q^2 - r_+^{2d-4})}{r_+^{2d-2}} = 0.
\end{align}
These two equations determine the charge and mass of the black hole in terms of the event horizon radius \(r_+\):
\begin{align}\label{eq: AdS d+1 lce}
    Q &\propto q = \left(r_+^{2d-4} + \frac{d}{d-2} r_+^{2d-2}\right)^{1/2}, \nonumber\\
    E &\propto M = 2 r_+^{d-2} + \frac{2(d-1)}{d-2} r_+^d.
\end{align}

One can demonstrate the AdS/EFT correspondence by equating the variables as follows:
\begin{align}
    r_+ = \sqrt{\frac{4(d-1)}{d} \lambda} u^{\frac{2}{d-2}},
\end{align}
such that the large charge expansion of the classical conformal field theory \eqref{eq: CFTd lce} matches that of the extremal black hole \eqref{eq: AdS d+1 lce}.

\bibliography{References.bib}{}

\end{document}